\documentclass[useAMS]{mn2e}
\usepackage{graphicx}
\usepackage{rotating}
\title[The SAURON Project - XX. The Spitzer 3.6 - 4.5 colour in early-type galaxies]{
The SAURON Project - XX. The Spitzer
[3.6] - [4.5]  colour in early-type galaxies: colours, colour gradients and inverted scaling relations}

\author[R.F.~Peletier, E.~ Kutdemir and SAURON]{Reynier F.~Peletier$^{1}$\thanks{E-mail: R.F.Peletier@astro.rug.nl}, Elif~
Kutdemir,$^{1,11}$ Guido van der Wolk,$^{1}$ Jes\'us Falc\'on-Barroso,$^{2,12}$ \newauthor 
Roland Bacon,$^{3}$ Martin Bureau,$^{4}$ , Michele Cappellari,$^{4}$ 
Roger L.\ Davies,$^{4}$ \newauthor 
P.~Tim de Zeeuw,$^{5,6}$ 
Eric Emsellem,$^{5,3}$ 
Davor Krajnovi\'c,$^{5}$  Harald Kuntschner,$^{5}$, \newauthor 
Richard M.\ McDermid,$^{7}$  
Marc Sarzi,$^{8}$ Nicholas Scott,$^{4}$ Kristen L. Shapiro,$^{9}$  \newauthor 
Remco C. E. van den Bosch,$^{10}$ and Glenn van de Ven$^{10}$ \\
$^{1}$Kapteyn Astronomical Institute, University of Groningen, Postbus 800, 
9700 AV Groningen, The Netherlands\\
$^{2}$Instituto de Astrof\'\i sica de Canarias, Via Lactea s/n, 38700 
La Laguna, Tenerife, Spain\\ 
$^{3}$Universit\'e Lyon 1, CRAL, Observatoire de Lyon, 9 avenue Charles Andr\'e, F-69230 Saint-Genis Laval, France\\
$^{4}$Sub-department of Astrophysics, University of Oxford, Denys Wilkinson Building, Keble Road, Oxford OX1 3RH, United Kingdom \\
$^{5}$European Southern Observatory, Karl-Schwarzschild-Str~2, 85748
Garching, Germany\\
$^{6}$Sterrewacht Leiden, Universiteit Leiden, Postbus 9513, 2300 RA Leiden, The Netherlands\\
$^{7}$Gemini Observatory, Northern Opertations Center, 670 N. Aohoku 
Place, Hilo, HI 96720, USA\\
$^{8}$Centre for Astrophysics Research, University of Hertfordshire, 
College Lane, Hatfield, Herts, AL10 9AB, UK\\
$^{9}$Aerospace Research Laboratories, Northrop Grumman Aerospace Systems,
Redondo Beach, CA 90278, USA\\
$^{10}$Max-Planck Institute for Astronomy, D-69117 Heidelberg, Germany\\
$^{11}$ T\"ubitak Uzay Space Technologies Research Institute, METU Campus, 06531, Ankara, Turkey\\
$^{12}$Departamento de Astrof\'isica, Universidad de La Laguna, E-38205 La Laguna, Tenerife, Spain\\
}

\begin{document}



\maketitle

\label{firstpage}

\begin{abstract}
We investigate the [3.6] - [4.5] Spitzer-IRAC colour behaviour of the early-type galaxies of the SAURON survey, a
representative sample of 48 nearby ellipticals and lenticulars. We investigate how this colour, which is unaffected by
dust extinction, can be used to constrain the stellar populations in these galaxies.  

We find a tight relation between the $[3.6]-[4.5]$ colour and effective velocity dispersion, a good mass-indicator in
early-type galaxies: ([3.6]-[4.5])$_e$ = (-0.109 $\pm$ 0.007) log $\sigma_e$ + (0.154 $\pm$ 0.016). Contrary to other colours in the optical and near-infrared, we  find that the colours become bluer
for larger galaxies. The relations are tighter when using the colour inside r$_e$  (scatter 0.013 mag) , rather than the much smaller r$_e$/8
aperture (scatter 0.023 mag) , due to the presence of young populations in the central regions.  We also obtain strong correlations  between
the $[3.6]-[4.5]$ colour and 3 strong absorption lines (H$\beta$, Mg~{\it b} and Fe 5015). Comparing our data with the
models of Marigo et al., which show that more metal rich galaxies are bluer, we can explain our  results in a way
consistent with results from the optical, by stating that larger galaxies are more metal rich. The blueing is caused by
a strong CO absorption band, whose line strength increases strongly with decreasing temperature and which covers a
considerable fraction of the 4.5 $\mu$m filter. In galaxies that contain a compact radio source, the $[3.6]-[4.5]$
colour is generally slightly redder (by 0.015 $\pm$ 0.007 mag using the $r_e/8$ aperture) than in the other galaxies, indicating small amounts of either hot dust,
non-thermal emission, or young stars near the center. 

We find that the large majority of the galaxies show redder colours with increasing radius.  Removing the regions with
evidence for young stellar populations (from the H$\beta$ absorption line)  and interpreting the colour gradients as
metallicity gradients, we find that our galaxies are more metal poor going outward. The radial [3.6] - [4.5] gradients
correlate very well with the metallicity gradients derived from optical line indices. We do not find any correlation
between the gradients and galaxy mass; at every mass galaxies display a real range in metallicity gradients. 

Consistent with our previous work on line indices, we find a tight relation between local [3.6] - [4.5] colour and 
local escape velocity. The small scatter from galaxy to galaxy, although not negligible, shows that the amount and
distribution of the dark matter relative to the visible light cannot be too different from galaxy to galaxy.

\end{abstract}

\newpage

%
%
\section{Introduction}
\label{sec:intro4}

Scaling relations play a fundamental role in the study of galaxy formation. As
an example, studies based on SDSS (Baldry et al. 2004) have shown that nearby
luminous galaxies exist in two varieties: galaxies on the red sequence, which
form an upper, red envelope in the relation between optical colour and
integrated absolute magnitude of the galaxies, and those that are found in the
blue cloud, that have bluer colours than those on the red sequence, and much
larger scatter. Galaxies on the red sequence are generally early-type galaxies,
while those in the blue cloud are generally late-type, disk galaxies that are 
blue because of their star formation, and will eventually move towards the red
sequence (Faber et al. 2007). 

The red sequence, or the colour-magnitude relation, is an example of a
mass-metallicity relation. On the relation one can see that as galaxies become
more massive and brighter they become redder in, e.g., $B-V$, indicating that
their metallicity increases. Since the red sequence consists of the reddest
galaxies at a certain magnitude, it is thought that it is the sequence of  the
oldest galaxies. Galaxies that are younger are found at bluer colours, towards
the blue cloud. The colour-magnitude relation in nearby clusters has been known
for more than 50 years. First established by Baum (1959), it was later used by
Sandage (1972) to derive the relative distance between the Virgo and Coma
clusters. Bower, Lucey \& Ellis (1992) showed that tight colour-magnitude
relations in $U-V$, $J-K$ and $V-K$ could be determined for those clusters.
Ellis et al. (1997) showed that a tight colour-magnitude relation continues to
exist in rich clusters up to z=1, with gradually fewer objects on the red
sequence as one goes to higher redshift (see also S\'anchez-Bl\'azquez et al.
2009).

Apart from studying the stellar populations of clusters of galaxies, the
colour-magnitude relation can also be used to study individual nearby galaxies.
If galaxies lie below the red sequence, they must be younger on average. This
technique has often been used in the literature (e.g. Schweizer \& Seitzer 1992,
Bower et al. 1992). A problem is that colours are affected by extinction, so
that features, such as central dust lanes, limit the accuracy of this method. To
circumvent this problem, one can study line strength - magnitude relations,
and,  instead of using magnitudes as a proxy of galaxy masses, one can obtain
smaller scatter using velocity dispersion ($\sigma$), in a certain galaxy
aperture (e.g., Terlevich et al. 1981, Bender, Burstein \& Faber 1993, Bernardi
et al. 2003, Kuntschner et al. 2006 (hereinafter Paper VI), S\'anchez-Bl\'azquez
et al. 2006).   Analysis of such line strength vs. velocity dispersion relations
using SSP models shows that the metallicity of a galaxy goes up with increasing
mass, and that more massive galaxies are older (e.g., Thomas et al. 2005, Kuntschner et al. 2010, Paper
XVII). 

This is Paper XX in the SAURON series, in which we study the properties of nearby galaxies using integral field spectroscopy and anciliary data. Here we study the 48 early-type galaxies of SAURON sample (de Zeeuw et al. 2002, Paper II)  which contains a representative range in absolute magnitude
and ellipticity, including elliptical and lenticular galaxies, in the Virgo
cluster and in the local field. For this sample, many physical parameters have
already been determined, making it easier to perform a thorough analysis of this
relation.

In Shapiro et al. 2010 (Paper XV)  we identify galaxies hosting low-level star
formation, as traced by PAH emission, using the Spitzer [3.6] - [8.0] colour,
with measured star formation rates that compare well to those estimated from
other tracers. Data from these satellites have revealed that star formation is
occurring in a significant fraction of early-type galaxies (Yi et al. 2005;
Young et al. 2009; Temi et al. 2009) and that this late-time star formation
contributes 1 - 10\% of the current stellar mass (Schawinski et al. 2007;
Kaviraj et al. 2007). From the SAURON maps, Sarzi et al. 2006 (Paper V)
identified a subset of the sample in which the low [OIII]/H$\beta$ emission-line ratios
and settled gas systems can only be interpreted as sites of star formation
activity. In these systems, the ionized gas is arranged in a regular, disk-like
configuration, with low velocity dispersion (Paper V). Stellar population
estimates (Paper VI; Paper XVII) and GALEX UV imaging (Jeong et al. 2009,
hereafter Paper XIII) have further revealed that these systems contain the young
stellar populations that must exist in the presence of on-going star formation.
Additional evidence for star formation in these galaxies comes from CO
observations, which reveal the presence of molecular gas and show that, in some
cases, this gas is organized into disks that are co-spatial and corotating with
the ionized gas and the young stars (Combes et al. 2007; Young et al. 2009).
However, it is important to note that despite this evidence for trace on-going
star formation in this subset of the sample, wide field optical imaging of the
SAURON galaxies continues to confirm that all the galaxies in this sample reside
solidly on the red sequence in optical colors (Falc\'on-Barroso et al. 2011, Paper
XIX). In the latter paper, the location of our galaxies in various scaling
relations is explored: the V-[3.6] -- magnitude and V-[3.6] -- $\sigma$
relations, the Fundamental Plane, as well as the Kormendy and Faber-Jackson
relations. It was found that the SAURON sample shows a tight colour -- $\sigma$ 
and Fundamental Plane relation, with an even smaller scatter for the  slow
rotators. Spirals also lie on these relations, albeit with more scattter.

As mentioned before, not only the age, but also the abundance distribution of various element changes as a function of galaxy mass.
Although the Mg~{\it b} line at 5170\AA\ has been used most often for line
strength -- $\sigma$ relations, other indices also mostly correlate with
velocity dispersion (e.g. S\'anchez-Bl\'azquez et al. 2006).  These correlations
depend on the individual abundance of the elements responsible  for the
absorption lines. In principle, it should be possible to derive the element
enrichment history of galaxies by studying various line strength - $\sigma$
relations. Results, however, are slow to come, mainly because very high quality
data are  required. For example, it has been known for years that [Mg/Fe]
increases for more massive galaxies. This is interpreted by assuming that Mg
(and other $\alpha$-elements) form more quickly, from SN type II (Peletier 1989,
Worthey et al. 1992). Some other elements follow Mg. For other elements, the
situation is much less clear. For example, the behaviour of the Ca II IR
triplet, which becomes less strong for more massive galaxies (Saglia et al.
2002, Cenarrro et al. 2003), is not well understood, since it is an
$\alpha$-element, which should follow elements like Mg, and show Ca/Fe values
larger than solar for large ellipticals. In reality, Ca seems to be 
under-abundant w.r.t. Fe (see e.g. Cenarro et al. 2003). In the  near-infrared,
even less is known, because of the lack of good data and models. In a recent
paper M\'armol-Queralt\'o et al. (2009) show indications that the index-$\sigma$
relation for the CO band at 2.3 $\mu$m is different in the Fornax cluster than
in the field, indicating that this band strength is not dependent on age and
metallicity in the same way as certain lines in the optical. It looks as if the 
CO-band contains a strong contribution from AGB stars of intermediate age.

The more one goes towards the near-infrared, the less the effects of extinction
are, and also the effects of young stars. In the infrared, the contribution from
old stars, and also evolved stars, such as those on the AGB, is dominant. In
principle, the only advantage in using line strengths instead of colours in the
near-IR is that line strengths are maybe simpler to understand. Accurate
colour-$\sigma$ relations, however, could significantly add to our understanding
of stellar populations in galaxies. In this paper, we decided to use the large,
high-quality database of Spitzer-IRAC images to study the stellar populations of
elliptical galaxies. Buta et al. (2010) nicely show that the 3.6 $\mu$m Spitzer
filter  in galaxies is dominated by stellar photospheric light. In early-type
galaxies, where the contribution from the interstellar medium is relatively
limited. The same can be said about the 4.5 $\mu$m filter. Comparing the light
in the two filters should give a colour which basically measures the average
temperature of the stars, which is determined by the stellar populations. This
study is particularly useful for interpreting the large surveys that are
currently being done with the {\it Spitzer Space Telescope} in its last phase,
with warm detectors. 

With the contribution from young stellar populations minimized, the Spitzer
[3.6]-[4.5] colour is well suited to measure metallicity gradients, which can
provide useful information to study galaxy formation. Classical monolithic
collapse scenarios  (Larson 1974, Carlberg 1984, Arimoto \& Yoshii 1987),as well
as their (revised) up-to-date versions  which start from semi-cosmological
initial conditions (e.g. Kawata 2001; Kobayashi 2004)  predict strong
metallicity gradients. In these scenarios primordial clouds of gas sink to the
centre of an overdensity where a rapid burst of star formation occurs. Infalling
gas mixes with enriched material freed from stars by stellar evolutionary
processes and forms a more metal rich population. Because the gas clearing time
(and hence the number of generations which enrich the interstellar medium) is
dependent on the depth of the potential well, the metallicity gradient is
dependent on the mass of the galaxy.

Mergers, dominant in hierarchical galaxy formation scenarios, will dilute
existing gradients (e.g. White 1980, di Matteo et al. 2009). The study of
metallicity gradients hence can help distinguish between competing scenarios of
galaxy formation and can eventually lead to a more detailed understanding of
those scenarios.  Many attempts have been made to model metallicity gradients in
more detail (Carlberg 1984, Arimoto \& Yoshii 1987, Chiosi \& Carraro 2002,
Kobayashi 2004, Pipino et al. 2008, 2010), of which the models by Kawata \&
Gibson (2003) are so far the most consistent with observations (Spolaor et al.
2009).

Much work has also been devoted to determine metallicity gradients in large
elliptical galaxies, either by use of colour gradients (Sandage 1972, Franx,
Illingworth \& Heckman 1989, Peletier et al. 1990, Peletier, Valentijn \&
Jameson 1990, Saglia et al. 2000, La Barbera et al. 2004, La Barbera et al.
2005, La Barbera \& de Carvalho 2009) or by spectroscopy (Davies, Sadler \&
Peletier 1993, Carollo \& Danziger 1993, {S{\'a}nchez-Bl{\'a}zquez}, Gorgas \&
Cardiel 2006,Ogando et al. 2008). In Paper XVII we have determined age,
metallicity and [$\alpha$/Fe] gradients for the SAURON early-type galaxy sample,
assuming that at every position in the galaxy the stellar populations could be
represented by SSPs.  There is consensus that colour and line strength
gradients, when converted to metallicity gradients, indicate that galaxies
become slowly more metal poor with increasing radius, however gradients are
shallower than predicted by classic monolithic collapse scenarios. Metallicity
gradients seem to peak around $\sigma$ = 100 - 150 km/s (Peletier et al. 1990, {S{\'a}nchez-Bl{\'a}zquez}, Gorgas \& Cardiel 2006,, Paper XVII).
Towards the faint end of the galaxy sequence gradients decrease (den Brok et al.
2011, Spolaor et al. 2010, Koleva et al. 2009). For galaxies with $\sigma ~<$
100 km/s there seems to be a strong correlation between colour gradient and
$\sigma$, with real scatter at every $\sigma$. For galaxies more massive than 
$\sigma ~=$ 150 km/s there does not seem to be any correlation between
metallicity gradient and $\sigma$ any more, and the scatter seems to be
considerable. To be able to use the information that is contained in the
metallicity gradients, it is important to understand their correlations, and
their scatter at a given $\sigma$. Since the Spitzer data at 3.6 and 4.5 $\mu$m
are relatively easy to interpret, without being affected by extinction,  and
with only a small dependence on the presence of young stellar populations, we
have calculated the gradients in the [3.6] - [4.5] colour, and have investigated
their properties.

The content of this paper is as follows: in Section 2 the [3.6] - [4.5] $\mu$m
colour and its use for stellar population analysis of galaxies is introduced. In
Section 3 the data reduction and analysis are described. In Section 4 we discuss
various scaling relations with the [3.6] - [4.5] colour. In section 5 the
gradients and their correlations are discussed.   The observational relations of
sections 4 and 5  are discussed in Section 6, after which in Section 7 the
conclusions are given.

\section{The $[3.6]-[4.5]$ colour as a tracer of old stellar populations}
\label{sec:colour}

As far as stellar populations of galaxies are concerned, the $[3.6]-[4.5]$ colour is almost virgin territory. Reasons for this are the fact that ground-based observations in the mid-infrared suffer from a large background from thermal emission from the Earth, and that there have been very few space missions with adequate instruments and
detectors up to now.  One of the first extragalactic mid-infrared spectra ever
published is the spectrum of M82 (Willner et al. 1977). Using single-detector
aperture photometry they found a featureless  Rayleigh-Jeans tail from 2
micron, combined with a strong Polyaromatic Hydrocarbon (PAH) feature at 3.3 $\mu$m. Much later, ISO-CVF
spectra of the elliptical galaxy NGC 1399  (Madden \& Vigroux
1999) showed that the spectrum between 5 and 10 $\mu$m is
consistent with a blackbody spectrum of T=4700K,  showing no evidence for
PAH-features. Kaneda et al. (2007) showed a high S/N spectrum of the early-type
galaxy NGC 1316, starting at 2.5 $\mu$m, which is almost featureless, apart from
an absorption feature at 4.5 $\mu$m. Very few other galaxy spectra are available
in the literature, also because  the IRS, the low resolution spectrograph on
board of the {\it Spitzer Space Telescope}, only covers the spectral region redward of
5 $\mu$m.

\begin{figure}
\begin{center}   
\includegraphics[width=8cm,clip]{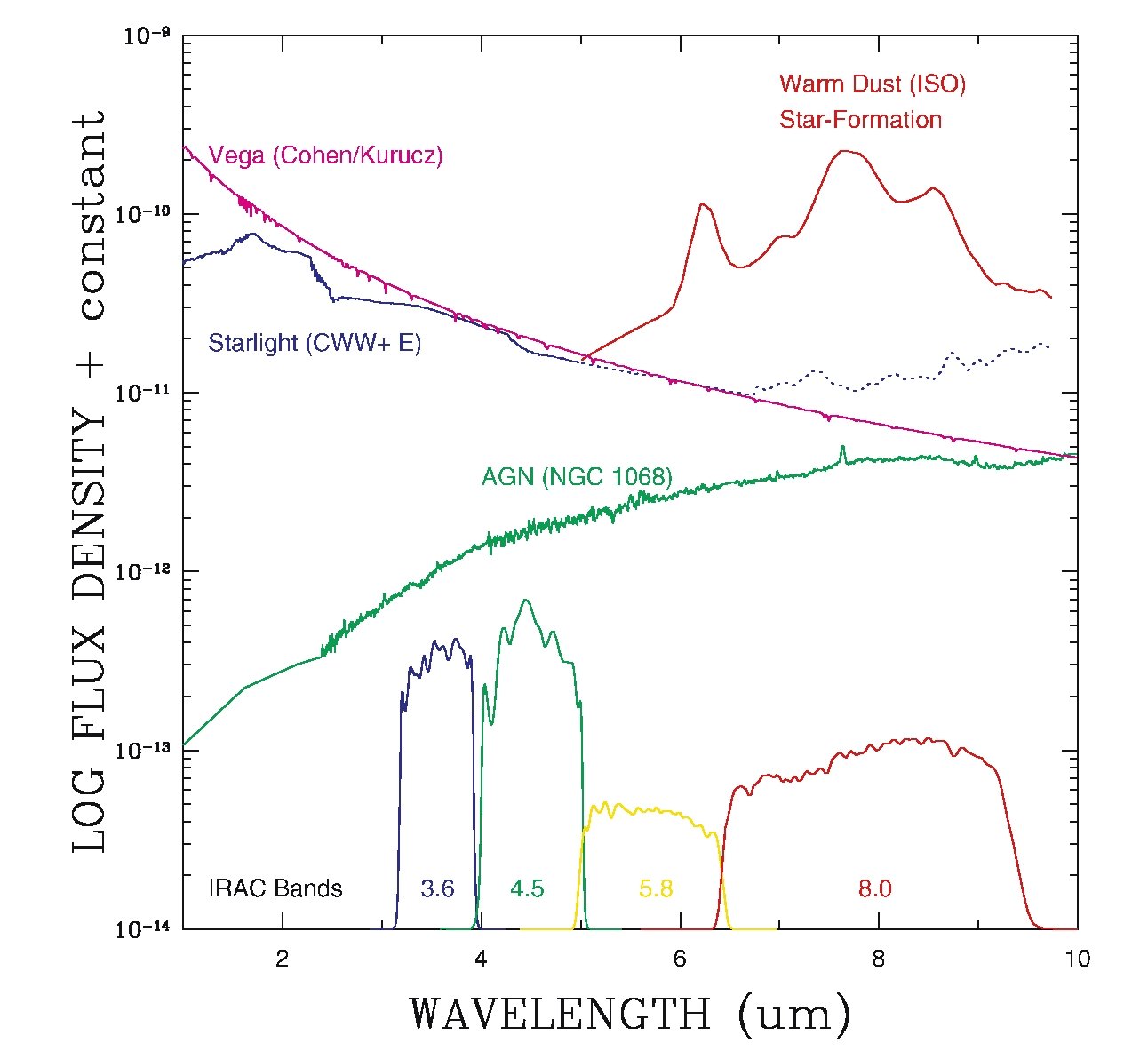}
 \caption{Spectral energy distributions of various galaxy components
(starlight, warm dust emitting in the PAH lines and active galactic nuclei)
together with the $3.6$ and $4.5\mu$m filters of IRAC (from Fazio 2005).}
\label{irac}
\end{center}
\end{figure}

Modelling of IRS spectra of elliptical galaxies shows that the emission mainly
comes from the Rayleigh-Jeans tail of stellar photospheric  spectra, together
with a silicate emission feature at 9-12 $\mu$m due to mass loss from AGB
stars. This feature has been used to determine the age of about 20 galaxies
(Bressan et al. 2006, Bregman et al. 2006), showing that all of these are old, a result implying that emission from young stars is very hard to find in the mid-IR. In a smaller fraction of these
galaxies line emission was found: unresolved line and silicate emission in M87
that likely originates in the dusty torus surrounding the AGN, and unresolved PAH emission in  NGC
4550 and NGC 4435 (Bressan et al. 2006). In Fig.~\ref{irac} the spectral energy
distributions of various galaxy components (starlight, warm dust emitting in the
PAH lines and active galactic nuclei) are shown together with the $3.6$ and
$4.5\mu$m filter transmission curves of IRAC on the {\it Spitzer Space Telescope} (Fazio 2005).

\begin{figure}
\centering
\includegraphics[width=9cm,clip]{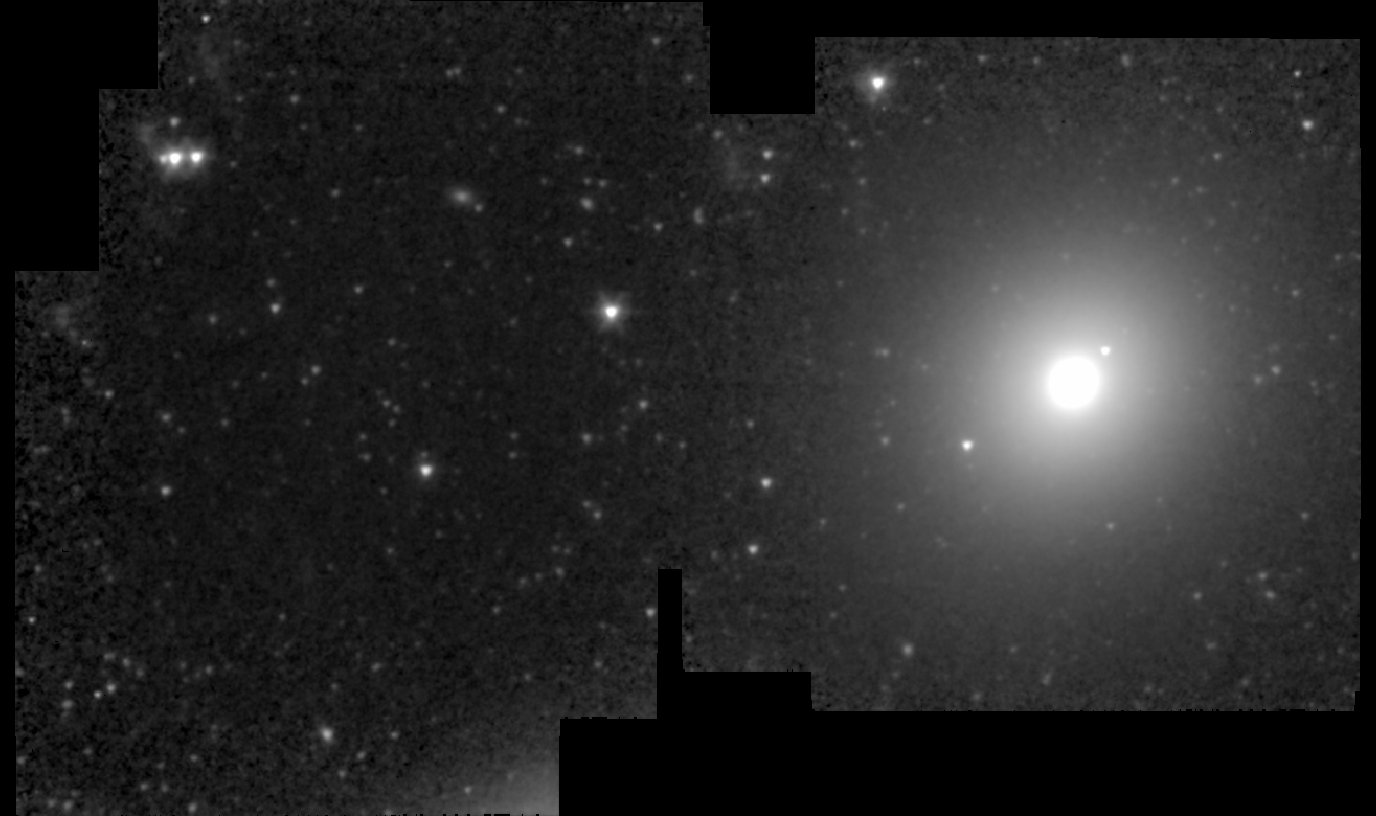}
\includegraphics[width=9cm,clip]{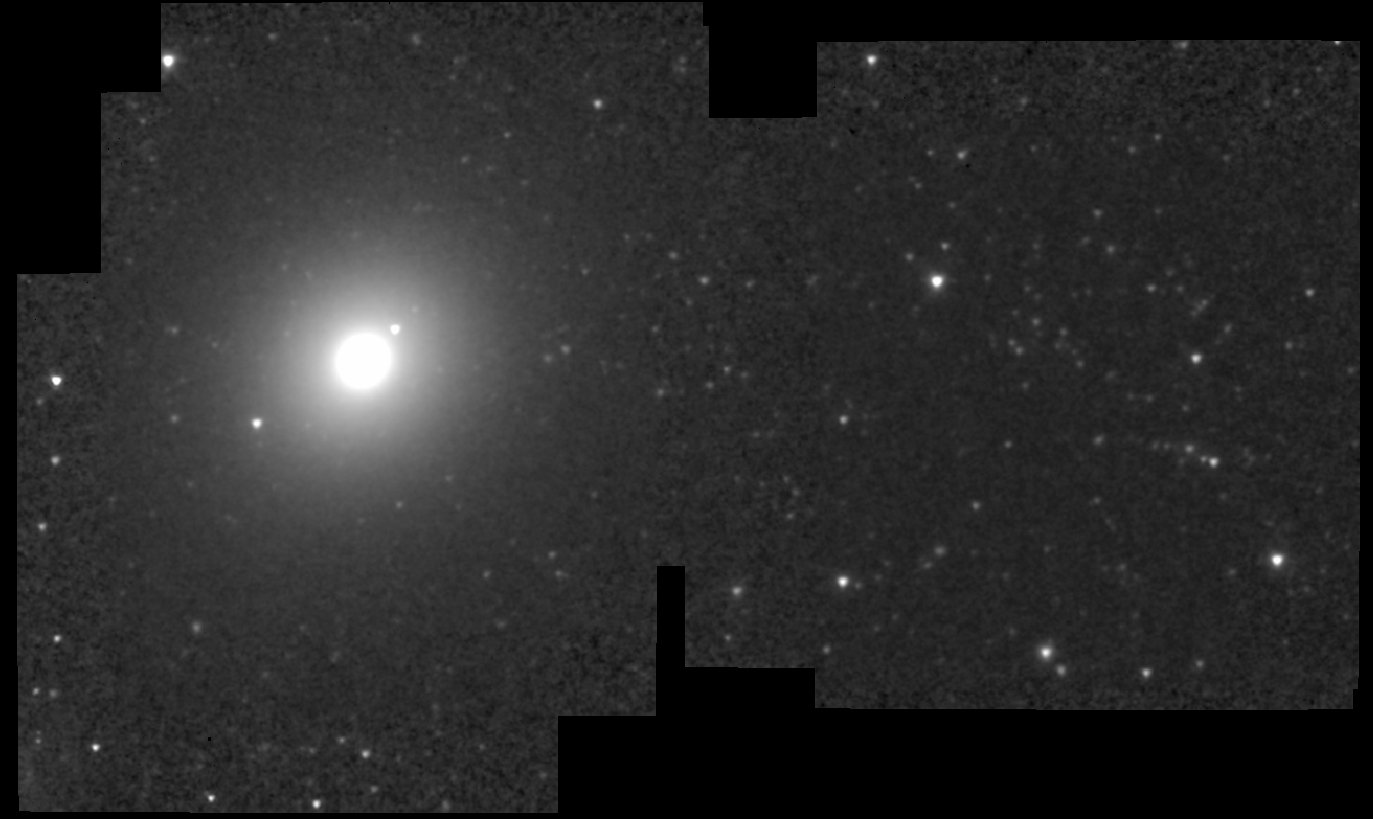}
\caption{Reduced Spitzer IRAC mosaics at 3.6 (top) and 4.5 $\mu$m of the elliptical
galaxy NGC 3379, showing the excellent quality of the data in both filters. The fact that
the galaxy looks so similar to the optical image suggests that most of the emission
is of stellar origin. The size of both images on the sky is 790$''$ $\times$ 487$''$.}
\label{n3379}
\end{figure}

\begin{figure}
\centering
\includegraphics[width=9cm,clip]{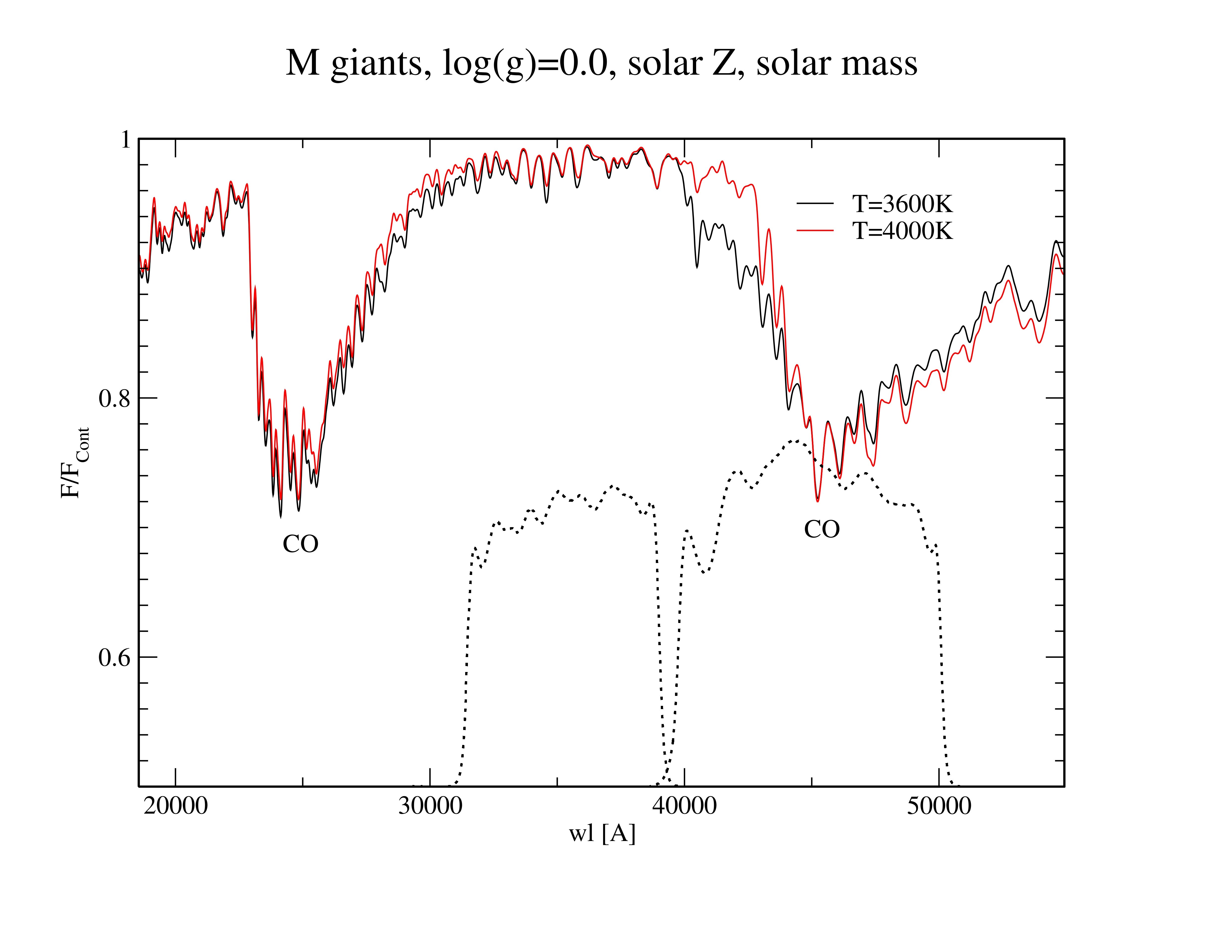}
\caption{Spectral energy distributions of two M-type giant stars from the
MARCS library (Bressan, private communication).  Two CO molecular absorption bands are visible,
the second one of which falls in the $4.5\mu$m filter of IRAC while the
$3.6\mu$m filter is free from molecular bands.  The figure shows that the
absorption feature is stronger for cooler stars and therefore that older stars have
bluer $[3.6]-[4.5]$ colours.The IRAC passbands are indicated in dashed lines. }
\label{mgiant}
\end{figure}

As far as extragalactic IRAC photometry is concerned, almost all the published
work is aimed at studying star formation and PAHs (e.g. Calzetti et al. 2005,
Paper XV, Mu\~noz-Mateos et al. 2009). 3.6 and 4.5 $\mu$m images
have been published, but the colour information is almost never quantitatively
analyzed.  This is remarkable, since the data quality is exceptional (see
Fig.\ref{n3379}). As an exception, Temi et al. (2008) show a [3.6] - [4.5]
$\mu$m colour profile, averaging a number of elliptical galaxies in the Spitzer
archive. They note that the profile is rather flat, but has a positive gradient,
i.e. that the galaxies become slightly redder in the outer parts. This behaviour
is noteworthy,  since it is contrary to the behaviour in all other optical filters,
in which elliptical galaxies generally become bluer outwards (e.g. Peletier et
al. 1990).  However, it can be understood when one realises that the 4.5
$\mu$m filter is dominated by a strong CO absorption band (e.g. Cohen et al.
1996). We show  some stellar model spectra in Fig.~\ref{mgiant} (Bressan, private communication), 
where the spectral energy distributions between $2$ and $5\mu$m
of two M-type giant stars with different temperatures are compared with each
other. When going radially outwards in an elliptical galaxy, the  metallicity
goes down. This causes an increase in the average stellar temperature, causing
the CO-band to become weaker, and the flux in the 4.5 $\mu$ filter to increase,
causing the  $[3.6] - [4.5]$ colour to redden, despite the increasing average
temperature. 

\section{Data reduction and analysis}
\label{sec:analysis}
\subsection{Spitzer data}

For our analysis we use pipeline-reduced $bcd$-images obtained with IRAC on the
{\it Spitzer Space Telescope} (Fazio et al. 2005) in the 3.6 and 4.5 $\mu$m filters. We
then used the standard MOPEX pipeline to make the final mosaics, sampling them
at a pixel size of 0.6$''$.  We used the zero magnitude flux
densities provided by Reach et al. (2005).

IRAC images suffer from complex scattering of light in the focal planes,
particularly for the Si:As detectors ($5.8$ and $8.0$ $\mu$m). This effect
causes sources to have extended halos, making it difficult to apply the flux
calibration obtained from stars to galaxies, and to apply accurate surface brightness profiles.  The scattering has two components:
1- the PSFs at $5.8$ and $8.0$ $\mu$m have more extended wings than expected, 2-
there is diffuse scattering which distributes a portion of the flux on a pixel
throughout the entire array.   The problem is discussed in detail at
http://irsa.ipac.caltech.edu/data/SPITZER/docs/irac/iracinstrumenthandbook/, Section 4.10 and 4.11. Here it is shown that the
effects of scattering are smallest in the  short-wavelength channels ($3.6$
and $4.5\mu$m) of IRAC, and that the aperture corrections in both filters are very similar.
Since the main aim of this work is the study of the
mid-IR colours, we cross-convolved each image with the PSF of the other filter,
using the PSFs and the IDL software of Gordon provided on the {\it Spitzer Space Telescope} website. This results in images with
exactly the same PSF at 3.6 and at 4.5 $\mu$m, except on scales larger than 10$''$, due to the limited size of the convolution kernels. Beyond this length scale it is most probable that the PSF in both images is very similar (Fig. 4.7 of the Spitzer handbook). We also decided not to apply any aperture corrections to our profiles, nor to our integrated colours, since these are rather uncertain. Similar
considerations led e.g. the SINGS team (Dale et al. 2009) to the same
conclusions. The tightness of the scaling relations shown in this paper shows that the PSF is sufficiently stable and constant from galaxy to galaxy that this approach is scientifically valid.  

\subsection{Colours and colour profiles}
\label{sec:aper}
The large field of view of SAURON (31$''$ $\times$ 40$''$, Paper II) made it possible to measure
velocity dispersions, morphological parameters such as boxiness and ellipticity, ionized
gas properties, line strength indices and GALEX UV fluxes for each galaxy  in
circular apertures of radius $r_e$ and $r_e/8$. We use the values tabulated in
Sarzi et al. (2006, Paper V), Paper VI, Emsellem et al. (2007, Paper IX), and
Bureau et al. (2011, Paper XVIII). Some important parameters are reproduced here
in Table~\ref{galpar}.  To investigate possible correlations with
these parameters, we measured the $[3.6]-[4.5]$ colour within the same
apertures. These numbers are are also given in Table~\ref{galpar}. In Fig.~\ref{n3379} 
example images of the galaxy NGC~3379 are given. The quality of these images is typical for the whole dataset. Technical details about the
observations, such as proposal ID and integration times, are given in Paper XV. 
The photon noise for all galaxies is so low that
Poisson errors in the magnitudes are negligible. We have calculated the errors
due to the uncertainties in the sky background, which are mainly due to small
structures in the background of the Spitzer images (see Table~\ref{galpar}).
Systematic errors due to, e.g., PSF scattering effects, might play a role,
but in the colours many systematic errors cancel out. From  the tightness in the
scaling relations presented in the next section it is clear that the systematic
errors in the colours  have to be very similar from galaxy to galaxy.


\onecolumn
\begin{table}
\caption{Galaxy parameters}
\label{galpar}
\small\centerline {
\begin{tabular}{crrrrcrccccccrr}
\hline
NGC   &  r$_e$ & $\sigma_e$ & $\epsilon_e$ & PA & (a4/a)$_e$ & ${\lambda_R}_e$ & $[3.6] - [4.5]_e$ & $\pm$ & $[3.6] - [4.5]_{{r_e}/8}$ & $\pm$ &
$\nabla([3.6]-[4.5])$ & $\pm$ & r$_{min}$ & r$_{max}$ \\
 & $''$ & km/s & ~ & deg & \multicolumn{1}{c}{\%} & ~ & mag & ~ & mag & ~ & mag/dex & ~ & \multicolumn{1}{c}{$''$} & \multicolumn{1}{c}{$''$} \\
$(1)$ &  $(2)$ &  $(3)$ & $(4)$ &  $(5)$ & $(6)$ & $(7)$ & (8) & (9) & (10) & (11) & (12) & (13) & (14) & (15)\\
\hline
~474 &        29    &	150  &      0.13&15  &       -0.14 &   0.200 &  -0.086 &0.008 & -0.108 &0.001 &   0.045&  0.016 &	 3.0&	   67.6 \\ 
~524 &        51    &	235  &      0.04&38  &       -0.16 &   0.278 &  -0.090 &0.003 & -0.099 &0.000 &   0.024&  0.009 &	 3.0&	  116.8 \\ 
~821 &        39    &	189  &      0.35&32  &       1.43  &   0.258 &  -0.089 &0.005 & -0.122 &0.000 &   0.075&  0.011 &	 3.0&	  105.4 \\ 
1023 &        48    &	182  &      0.36&82  &       0.54  &   0.385 &  -0.096 &0.003 & -0.113 &0.000 &   0.039&  0.007 &	 5.0&	  182.7 \\ 
2549 &        20    &	145  &      0.49&0   &       2.86  &   0.539 &  -0.097 &0.001 & -0.132 &0.001 &   0.064&  0.010 &	 3.0&	  117.5 \\ 
2685 &        20    &	96   &      0.59&36  &       2.93  &   0.716 &  -0.042 &0.004 & -0.067 &0.001 &   0.021&  0.009 &	 3.0&	   51.8 \\ 
2695 &        21    &	188  &      0.21&170 &       0.36  &   0.561 &  -0.086 &0.004 & -0.125 &0.001 &   0.037&  0.015 &	 3.0&	   60.1 \\ 
2699 &        14    &	124  &      0.19&50  &       1.04  &   0.450 &  -0.097 &0.003 & -0.120 &0.001 &   0.081&  0.020 &	 3.0&	   43.9 \\ 
2768 &        71    &	216  &      0.46&94    &     0.12  &   0.268 &  -0.073 &0.009 & -0.090 &0.000 &   0.027&  0.011 &	 6.0&	  190.2 \\ 
2974 &        24    &	233  &      0.37&42  &       0.64  &   0.602 &  -0.089 &0.003 & -0.057 &0.001 &   0.032&  0.017 &	 7.0&	  117.8 \\ 
3032 &        17    &	90   &      0.17&95  &       0.44  &   0.416 &  -0.020 &0.007 & -0.045 &0.001 &  -0.007&  0.071 &	14.9&	   32.3 \\ 
3156 &        25    &	65   &      0.47&48  &       -0.04 &   0.713 &  -0.045 &0.006 & -0.054 &0.001 &   0.013&  0.013 &	 3.0&	   42.0 \\ 
3377 &        38    &	138  &      0.50&42  &       0.94  &   0.475 &  -0.057 &0.005 & -0.098 &0.001 &   0.059&  0.010 &	 3.0&	  134.7 \\ 
3379 &        42    &	201  &      0.11&71  &       0.16  &   0.145 &  -0.101 &0.002 & -0.124 &0.000 &   0.026&  0.008 &	 3.0&	  178.4 \\ 
3384 &        27    &	145  &      0.20&50  &       1.13  &   0.414 &  -0.081 &0.002 & -0.071 &0.000 &  -0.006&  0.014 &	13.5&	  114.1 \\ 
3414 &        33    &	205  &      0.23&2   &       1.80  &   0.062 &  -0.090 &0.005 & -0.136 &0.001 &   0.078&  0.013 &	 3.0&	   86.4 \\ 
3489 &        19    &	98   &      0.29&68  &       -0.61 &   0.602 &  -0.064 &0.002 & -0.094 &0.001 &   0.038&  0.008 &	 3.0&	   95.4 \\ 
3608 &        41    &	178  &      0.20&80  &       -0.21 &   0.038 &  -0.084 &0.006 & -0.091 &0.000 &   0.031&  0.018 &	 6.0&	   96.1 \\ 
4150 &        15    &	77   &      0.28&145 &       -0.32 &   0.584 &  -0.040 &0.003 & -0.034 &0.001 &   0.038&  0.028 &	 9.0&	   63.0 \\ 
4262 &        10    &	172  &      0.11&153 &       1.28  &   0.245 &  -0.103 &0.001 & -0.131 &0.000 &   0.086&  0.014 &	 3.0&	   50.6 \\ 
4270 &        18    &	122  &      0.44&106 &       -0.64 &   0.446 &  -0.073 &0.004 & -0.085 &0.001 &  -0.000&  0.009 &	 3.0&	   40.0 \\ 
4278 &        32    &	231  &      0.13&34  &       -0.15 &   0.149 &  -0.088 &0.002 & -0.109 &0.000 &   0.039&  0.010 &	 3.0&	  107.8 \\ 
4374 &        71    &	278  &      0.13&126 &       -0.40 &   0.023 &  -0.098 &0.004 & -0.116 &0.000 &   0.023&  0.008 &	 3.0&	  191.4 \\ 
4382 &        67    &	196  &      0.22&19  &       0.59  &   0.155 &  -0.090 &0.007 & -0.099 &0.001 &   0.011&  0.006 &	 3.0&	  158.0 \\ 
4387 &        17    &	98   &      0.32&143 &       -0.76 &   0.408 &  -0.083 &0.003 & -0.093 &0.001 &   0.034&  0.013 &	 3.0&	   48.6 \\ 
4458 &        27    &	85   &      0.14&4     &     0.41  &   0.046 &  -0.060 &0.007 & -0.086 &0.001 &   0.055&  0.018 &	 3.0&	   49.7 \\ 
4459 &        38    &	168  &      0.17&106 &       0.22  &   0.436 &  -0.089 &0.004 & -0.096 &0.000 &   0.004&  0.008 &	 3.0&	   70.0 \\ 
4473 &        27    &	192  &      0.43&92  &       1.03  &   0.195 &  -0.099 &0.002 & -0.120 &0.000 &   0.036&  0.009 &	 3.0&	  137.5 \\ 
4477 &        47    &	162  &      0.23&77  &       2.04  &   0.215 &  -0.080 &0.006 & -0.099 &0.001 &   0.036&  0.009 &	 3.0&	  101.1 \\ 
4486 &        105   &	298  &      0.07&159 &       -0.07 &   0.019 &  -0.106 &0.006 & -0.090 &0.001 &   0.020&  0.009 &	 7.3&	  157.0 \\ 
4526 &        40    &	222  &      0.41&112 &       -1.92 &   0.476 &  -0.081 &0.002 & -0.052 &0.000 &   0.014&  0.015 &	14.3&	  214.4 \\ 
4546 &        22    &	194  &      0.36&79  &       0.69  &   0.604 &  -0.096 &0.002 & -0.131 &0.000 &   0.068&  0.008 &	 3.0&	  111.2 \\ 
4550 &        14    &	110  &      0.62&179 &       2.36  &   0.091 &  -0.067 &0.002 & -0.076 &0.001 &   0.039&  0.010 &	 3.0&	   67.0 \\ 
4552 &        32    &	252  &      0.06&125 &       0.00  &   0.049 &  -0.107 &0.002 & -0.137 &0.000 &   0.062&  0.010 &	 3.0&	  173.8 \\ 
4564 &        21    &	155  &      0.43&48  &       1.33  &   0.586 &  -0.110 &0.002 & -0.151 &0.000 &   0.057&  0.009 &	 3.0&	   82.0 \\ 
4570 &        14    &	173  &      0.44&158 &       1.90  &   0.561 &  -0.115 &0.001 & -0.138 &0.001 &   0.060&  0.008 &	 3.0&	  118.9 \\ 
4621 &        46    &	211  &      0.35&163 &       1.66  &   0.268 &  -0.093 &0.004 & -0.129 &0.000 &   0.059&  0.009 &	 3.0&	  177.1 \\ 
4660 &        11    &	185  &      0.41&100 &       0.66  &   0.472 &  -0.101 &0.001 & -0.127 &0.000 &   0.065&  0.011 &	 3.0&	   60.2 \\ 
5198 &        25    &	179  &      0.14&26  &       -0.17 &   0.060 &  -0.084 &0.004 & -0.106 &0.001 &   0.044&  0.023 &	 5.6&	   58.2 \\ 
5308 &        10    &	208  &      0.53&59  &       4.74  &   0.483 &  -0.097 &0.001 & -0.142 &0.000 &   0.030&  0.008 &	 3.0&	   85.8 \\ 
5813 &        52    &	230  &      0.17&136 &       -0.03 &   0.063 &  -0.091 &0.009 & -0.106 &0.000 &   0.013&  0.011 &	 3.0&	  144.2 \\ 
5831 &        35    &	151  &      0.20&141   &     0.46  &   0.049 &  -0.080 &0.006 & -0.120 &0.001 &   0.063&  0.015 &	 3.0&	   74.7 \\ 
5838 &        23    &	240  &      0.28&41  &       0.34  &   0.518 &  -0.098 &0.002 & -0.109 &0.000 &   0.020&  0.013 &	 5.4&	  120.0 \\ 
5845 &        4     &	239  &      0.31&140 &       0.63  &   0.357 &  -0.115 &0.000 & -0.102 &0.000 &   0.076&  0.017 &	 3.0&	   30.0 \\ 
5846 &        81    &	238  &      0.07&58  &       -0.38 &   0.024 &  -0.103 &0.010 & -0.123 &0.001 &   0.034&  0.010 &	 3.0&	  190.4 \\ 
5982 &        27    &	229  &      0.28&106 &       -0.92 &   0.093 &  -0.090 &0.004 & -0.118 &0.000 &   0.048&  0.013 &	 3.0&	  103.6 \\ 
7332 &        11    &	125  &      0.39&155 &       1.35  &   0.390 &  -0.086 &0.001 & -0.135 &0.000 &   0.044&  0.007 &	 3.0&	   80.0 \\ 
7457 &        65    &	78   &      0.43&125 &       0.20  &   0.570 &  -0.051 &0.008 & -0.061 &0.001 &   0.008&  0.010 &	 3.0&	  107.1 \\ 
\hline	    
\end{tabular}}
\medskip
Column~(1): object ID (NGC number);
(2): Effective radius (from Paper XIX). ; (3): Integrated velocity dispersion inside the effective radius
($\sigma_e$), from Paper XIX; (4): Mean ellipticity within r$_e$, from Paper IX; (5): Major axis
position angle from N through E; (6): Mean isophote shape parameter 
a4/a (in percent, from Paper IX); (7): $\lambda_R$ within
r$_e$ , from Paper IX; (8): Integrated $[3.6]-[4.5]$ colour in circular apertures of r=r$_e$; 
(9): Error in this quantity due to sky background errors alone; (10): Integrated $[3.6]-[4.5]$ colour in circular apertures of r=r$_e$/8; 
(11): Error in this quantity due to sky background errors alone.
The units in columns (8)-(11) are Vega magnitudes calculated using zero magnitude flux densities
provided by Reach et al. (2005); (12): colour gradient $\Delta([3.6]-[4.5])/\Delta(\log~r)$ per dex in radius; (13) error in (2); (14) and (15): 
minimum and maximum of the radial major axis range in which the colour gradient has been determined.
\label{galpars}
\end{table}

\twocolumn

\begin{table}
\caption{Colour profiles}
\label{tab_prof}
\small\centerline {
\begin{tabular}{lccc}
\hline
Galaxy & r$_{maj}$ & [3.6] -- [4.5] & $\pm$ \\
~ & arcsec & mag arcsec$^{-2}$ & ~ \\
\hline
NGC ~474  &    1.01  &  -0.116  &   0.001 \\
NGC ~474  &    1.34  &  -0.114  &   0.001 \\
NGC ~474  &    1.77  &  -0.112  &   0.001 \\
NGC ~474  &    2.35  &  -0.111  &   0.001 \\
\hline	    
\end{tabular}}
\medskip
Column~(1): Galaxy ID; (2): major axis radius in arcsec; (3): $[3.6]-[4.5]$ colour;
(4): Error in this colour. The whole table is available in electronic form.
\end{table}

\begin{figure}
   \centering
   \includegraphics[height=\textheight]{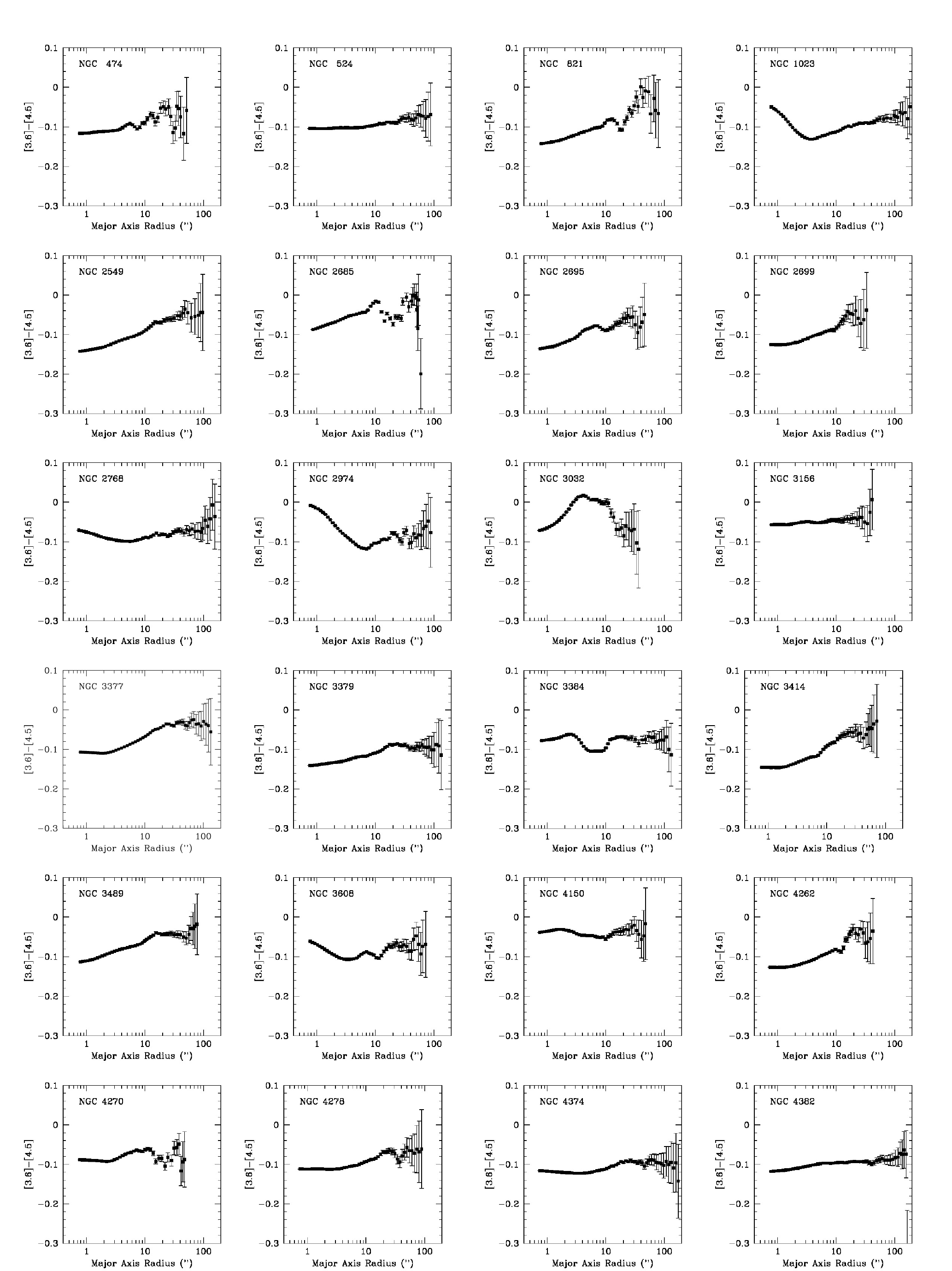}
 \caption{a) $[3.6]-[4.5]$  colour profiles in mag~arcsec$^{-2}$  of the sample galaxies.} 
         \label{prf1}
   \end{figure}

We proceeded by obtaining the [3.6] and [4.5] surface brightness profile of each
galaxy  using GALPHOT (J\o rgensen et al. 1992), implemented in IRAF. To do this, we masked
the foreground stars and instrumental features in the mosaics, such as high or
low columns and rows. In this process the galaxy center was chosen to be at the
3.6 $\mu$m intensity peak, which was always the same, within the errors, as the
intensity peak at 4.5 $\mu$m.   We first fitted concentric ellipses on the 3.6
$\mu$m images,  allowing ellipticity and major axis position angle to vary at
every radius, for each galaxy, to find the best fit, while keeping the center
fixed. We then determined a unique representative ellipticity and position
angle, which  we fixed (see Table \ref{galpar}), and then determined surface
brightness profiles in [3.6] and [4.5] on these isophotes. In Fig.~\ref{prf1}a
and \ref{prf2}b, we give the $[3.6]-[4.5]$ colour profiles. One can see that
there is no colour information in the inner 1$''$, since the PSFs of the two
filters were homogenised. Generally we see that the galaxy colours are becoming
redder when going outwards, opposite to e.g. colours in the optical and
near-infrared. On top of that we sometimes see features  near the center,
producing colours that are slightly redder than expected from extrapolating the
outer regions inwards. The colour profiles are tabulated in Table
\ref{tab_prof}.

\onecolumn

\addtocounter{figure}{-1}

  \begin{figure}
   \centering
   \includegraphics[height=\textheight]{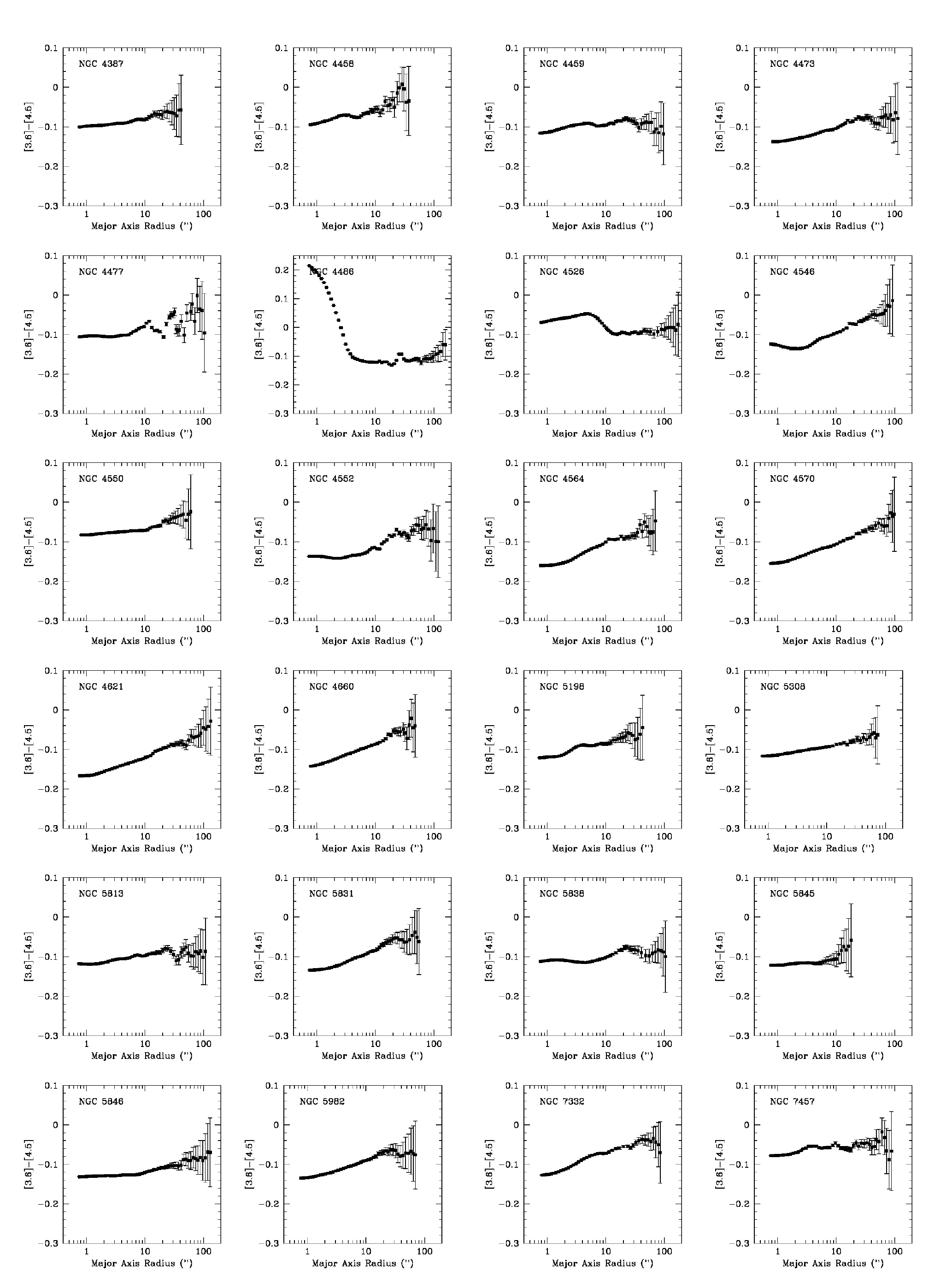}
\caption{b) $[3.6]-[4.5]$  colour profiles  in mag~arcsec$^{-2}$ of the sample galaxies, continued.} 
\label{prf2}
\end{figure}
\twocolumn

\subsection{Colour gradients}
\label{sec:gradients}

Early-type galaxies, because of their classification as red objects, mainly
consist of relatively old stellar populations, with limited areas that also
contain younger stars. The younger stars in our sample are generally found in
central rings or disks (Paper XVII). Further out, the colours of early-type
galaxies are rather uniform,  indicating that the stellar populations there are
generally old, and probably  only varying because of differences in metallicity
(see e.g. Peletier et al. 1990). In this subsection we try to quantify the
stellar population gradients in the old stellar populations. To do this, we 
select the regions  dominated by the old stellar populations using the H$\beta$
maps of Kuntschner et al. (2006), and use them to define the radial regions in
the radial colour profiles that we use to fit the radial colour  gradients of
the old stellar populations. In this way we do not include features like the
central disk in NGC~4526 (Fig. ~\ref{n4526}). Near the center of the galaxies we
don't use the data inside 3$''$, since there is very little colour information
there as a result of the coarse sampling of IRAC. All data points  are used for
which the errors in the colour due to sky background uncertainties are less than
0.1 mag.   The ranges are given in Table~\ref{galpar}. In the radial regions that
we use the [3.6] - [4.5] colour generally shows a linear behaviour as a function
of logarithmic radius. Because of this, we fit the logarithmic slope
$\Delta([3.6]-[4.5])/\Delta(\log~r)$, where the radii are major axis radii,
weighting the points with their sky background errors.   We find that most
profiles show a positive gradient: they are redder in the outer parts, contrary
to all optical and optical-near infrared filters, in which galaxies are bluer in
the outer regions.  As explained in Section \ref{sec:aper}, this corresponds to
decreasing metallicities and increasing ages when going outwards radially.

\subsection{Stellar population models}
\label{models}

\begin{figure}
\centering
\includegraphics[width=9cm,clip]{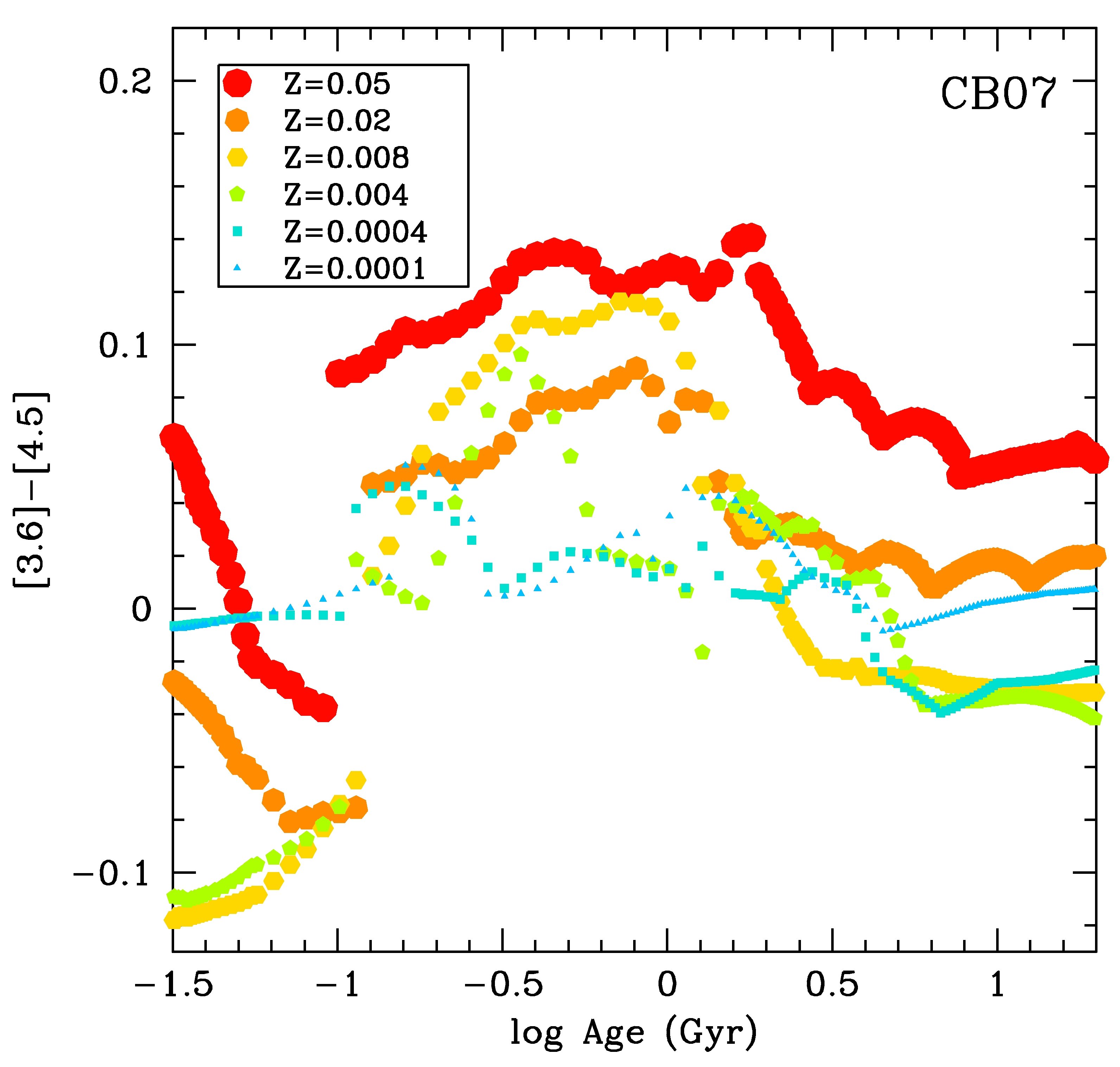}
\includegraphics[width=9cm,clip]{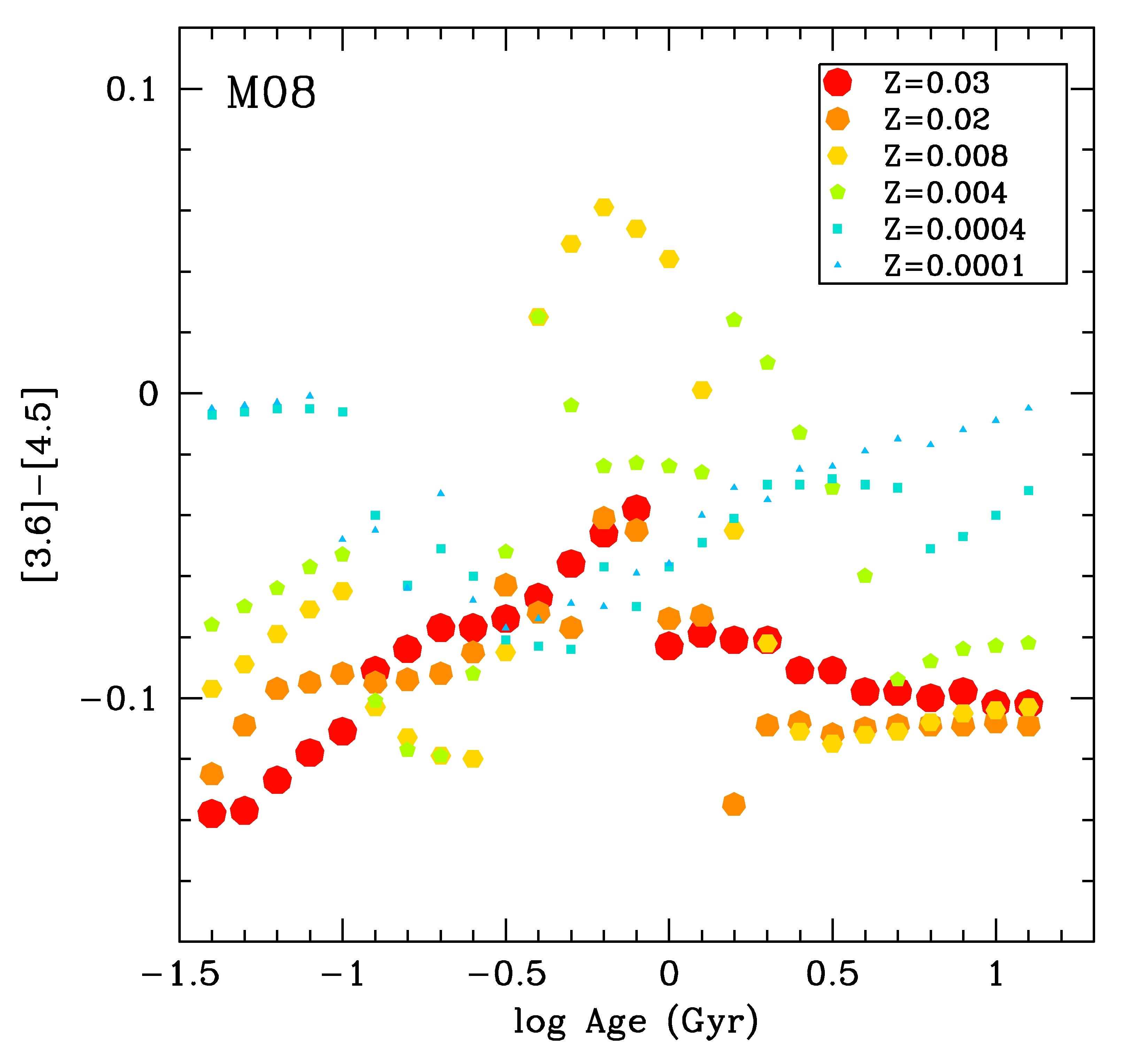}
\caption{[3.6]-[4.5] $\mu$m colours in mag~arcsec$^{-2}$  of 2 sets of models. Top: SSP Models of 
Charlot \& Bruzual (2007) as a function of age and metallicity.
\textit{Bottom}: SSP models of Marigo et al. (2008) as a function of age and
metallicity  (Here some tracks at Z=0.0001 are from Bertelli et al. (1994)).} 
\label{model}
\end{figure}

Very few stellar population models are available that try to reproduce the
behaviour of galaxies in the Spitzer-IRAC filters. This is due partly to the lack
of observational spectra that can be used to test these models, and partly
because of the difficulty to model the late stages of stellar evolution,
which are the dominant stars at 3.6 and 4.5 $\mu$m. Here we take a look at the
models of  Charlot \& Bruzual (2007, unpublished) and Marigo et al. (2008).  Both models are not independent, since Charlot \& Bruzual follow the TP-AGB prescription of Marigo et al.  In Fig.~\ref{model} we plot the
$[3.6]-[4.5]$  colours of the single age, single metallicity stellar population
models of  Charlot \& Bruzual (2007) and  Marigo et al. (2008). 
The symbols in both figures are the same,
except for the highest metallicity (Z=0.05 for Bruzual \& Charlot (2003) and Z=0.03 for
Marigo et al. (2008)).

In the models of
Charlot \& Bruzual (2007) the $[3.6] - [4.5]$ colour becomes redder with increasing
metallicity. In Marigo et al. (2008), this behaviour is not seen, and here a much bluer $[3.6] - [4.5]$ colour is predicted for
the highest 3 metallicities, probably because of the inclusion of the CO-band in the 4.5
$\mu$m filter in their models. 
For model ages above about 2 Gyr there is almost no
metallicity-dependence at Z=0.008 or higher (the 
regime of our observations, Paper XVII). 
For younger ages, the $[3.6] - [4.5]$ colour starts to
fluctuate considerably as a function of age and metallicity, in both models, since this is the regime in which AGB stars are
prominent. A galaxy population can become bluer by increasing its metallicity,
or by adding young ($10^7$ -- $10^8$ year old) AGB stars. The population at 1-2 Gyr is particularly red. 
   
\section{Global colour relations}
\label{sec:relations}

In this section we explore relations between the [3.6] -- [4.5] colour and
various mass indicators of galaxies in the SAURON sample. We try to find out
what the effect of young stellar populations on the [3.6] -- [4.5] colour is and
how it is affected by metalllicity. In earlier SAURON papers  a subset of the
sample was identified in which young stellar populations must exist in the
presence of on-going star formation, from [OIII]/H$\beta$ emission line ratios,
[3.6] -- [8.0] Spitzer maps, GALEX UV colours, molecular CO observations  and
optical line strength indices. It was found that these star-forming galaxies in
the SAURON sample are the minority of the population,  with the exact fraction
depending on the tracer used (5/48: from [OIII]/H$\beta$ emission line ratios,
Paper V; 9/34: from UV emission, Paper XIII). Of the 48 SAURON galaxies, roughly
half have 8.0$\mu$m non-stellar maps with significant power. In 8 of the 48
SAURON galaxies, there is strong evidence that the 8.0$\mu$m non-stellar flux
reflects PAH transitions excited by young stars. For these the 8.0$\mu$m maps
are consistent with those of H$\beta$ absorption line strength and the blue
GALEX colors. Mostly the star formation is centrally concentrated.

In Falc\'on-Barroso et al. (2011) the V-[3.6] -- magnitude and V-[3.6] --
$\sigma$ relations are studied. Here it was found that the SAURON sample shows a
tight colour -- $\sigma$ relation, with an even smaller scatter for the  slow
rotators. Spirals also lie on these relations, albeit with more scattter. The
colour -- $\sigma$ relation is relatively hard to analyze, since the V-[3.6]
colour is affected by both young stellar populations  and dust extinction. 
Since dust and young stellar populations are often found together, objects in
the SAURON sample with wide spread star formation (e.g., NGC 3032, NGC3156,
NGC4150, NGC3489) do not display the  bluest colour for a given velocity
dispersion. Here we analyze the [3.6]-[4.5] colour, which is little affected
by extinction, and which for that reason might be easier to interpret.

\subsection{The [3.6]-[4.5] -- $\sigma$ relation}

In the past, the colour-magnitude relation, and later also the colour-$\sigma$ relation have played an important role to investigate the nature of early-type galaxies. Here the central velocity dispersion and total magnitude are used as proxies of
galaxy mass, and the colour as a stellar population indicator. We also use the total stellar mass, from Jeans modelling (Scott et al., Paper XIV). In rough terms, it is
believed that along the relation, on the red sequence,
the mass of galaxies increases, possibly with slight changes in average age, while deviations from the relation are caused
by the presence of young stellar populations, extinction by dust, and possibly
objects that are dynamically young, e.g. merger remnants.

\begin{figure}
\begin{center}
\includegraphics[width=\textwidth]{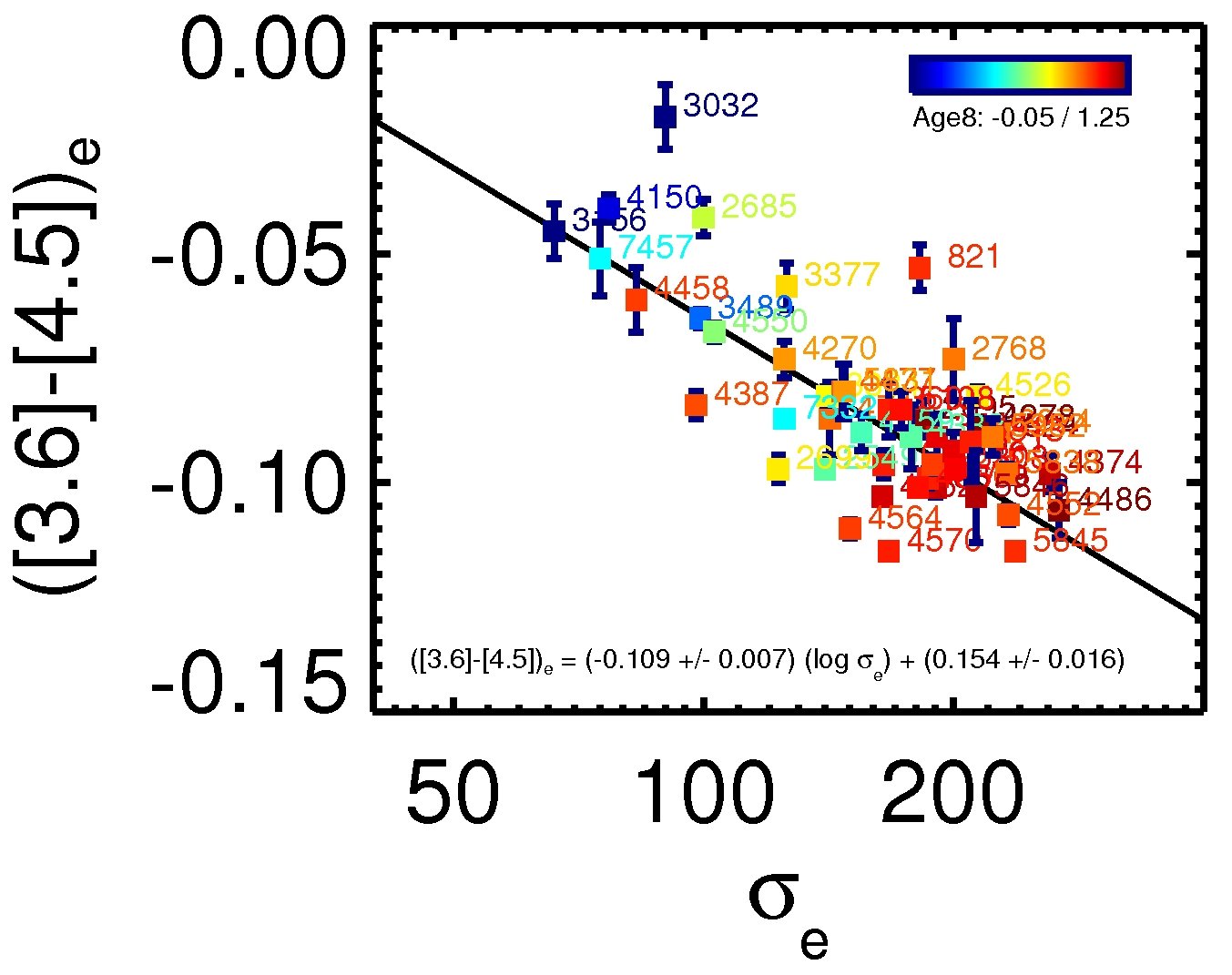}
\caption{ $[3.6]-[4.5]$  colour as a function of velocity dispersion  in
km/s. The velocity dispersion has been measured within r$_e$. Here the 
colour, determined within 1 effective radius, is shown, while
the points have been colour coded according to the age inside r$_e$/8} 
\label{mass_met}	 
\end{center}
\end{figure}

\begin{figure}
\begin{center}
\includegraphics{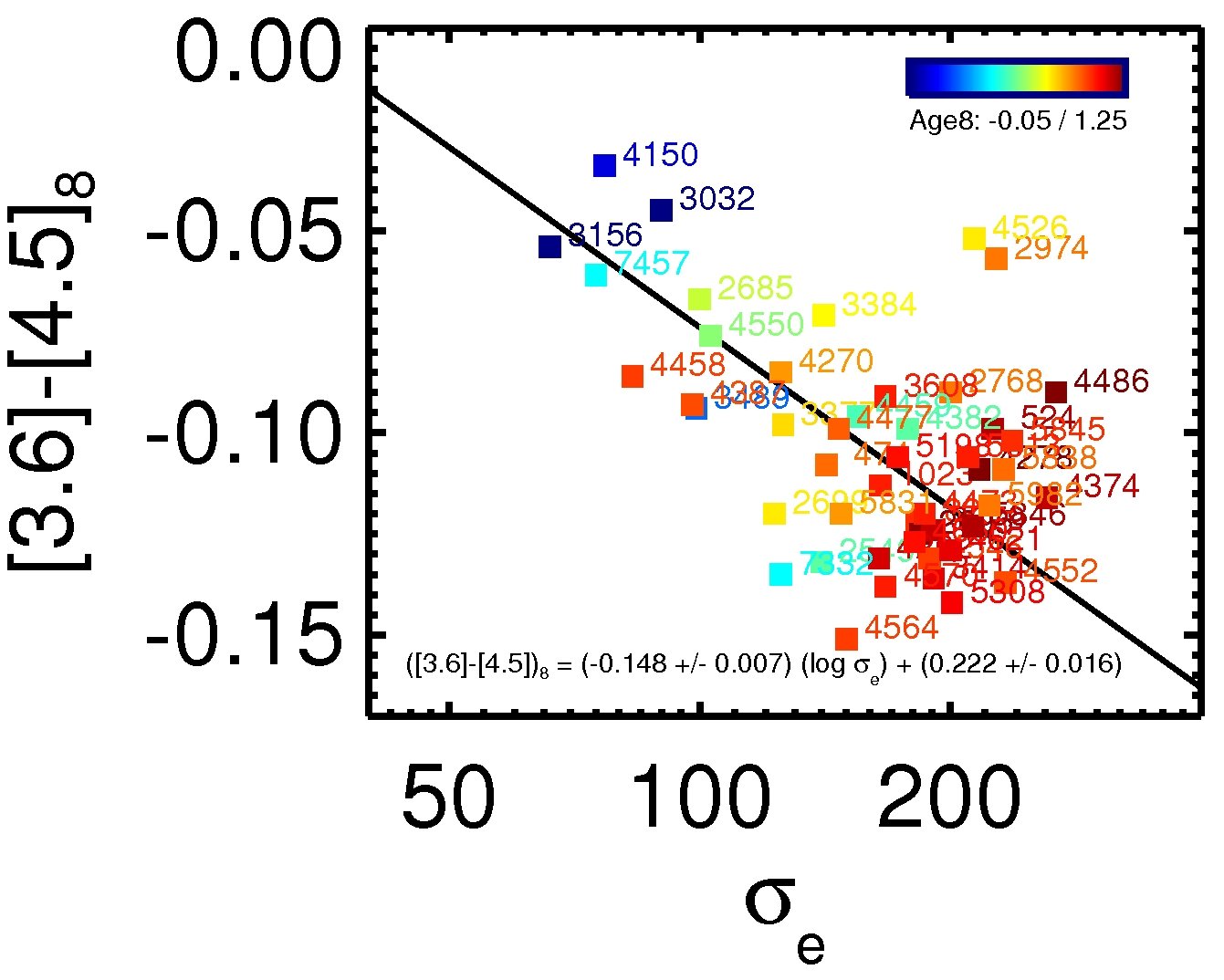}
\caption{Same as Fig.~\ref{mass_met}, except that
here the $[3.6]-[4.5]$ colour within r$_e$/8 is shown.} 
\label{mass_met8}	 
\end{center}
\end{figure}

Since in other papers of the SAURON series many quantities have been calculated
in circular apertures of radius $r_e/8$ and $r_e$, we have used these apertures
throughout the paper. In Fig.\ref{mass_met} and \ref{mass_met8} we show the $[3.6]-[4.5]$ vs.
$\sigma$ relation for these two apertures. In Fig.~\ref{mass_met} the colour inside $r_e$ is
given, while in Fig.~\ref{mass_met8} the $r_e/8$ aperture is used. For the velocity
dispersion, we always use the value within $r_e$ in this  paper, since this
gives the least scatter. The points have been coloured with the SSP-age within
$r_e/8$ (from Paper XVII). Had we used ages within $r_e$, the figures would have
been similar, but with a smaller range in colours. This is because in these
early-type galaxies many more young features are seen in the inner parts than
further out (Paper XVII, Paper V, Paper VI, Paper XIII). In
Fig.~\ref{mass_met} and Fig.~\ref{mass_met8} we have plotted the error bars in the colours that are due to
uncertainties in the sky background.  In Fig.~\ref{mass_met8} these errors are much smaller than the
dispersion in the points. The figures also show linear fits obtained using the least squares
method. All linear fits in this paper have been obtained using the 
program FITEXY from the IDL Astro-library (Landsman 1993) has been used, with errors 
in both variables. The errors in $\sigma_e$ have been taken to be 5\% (Paper XIX), and 
the colours a photometric uncertainty of 0.005 mag has been added quadratically to the sky background 
errors described above.  
The fit shows that there is a
strong dependence of the $[3.6]-[4.5]$ $\mu$m colour on the velocity dispersion
in both apertures, with an  additional RMS scatter, not explained by observational uncertainties,
of 0.013 mag for the $r_e$ aperture and
0.023 mag for r=$r_e/8$.

The colour-$\sigma$ relation shows that more massive galaxies are bluer. The
colour coding shows that these galaxies are at the same time older, if one
considers the luminosity-weighted SSP-ages. We see that[3.6]-[4.5] 
within $r_e/8$ is on the average 0.02 mag bluer than within $r_e$,
something that is due to colour
gradients within the galaxies. We find that the scatter within $r_e/8$ is
somewhat larger than within $r_e$. At first view the interpretation seems 
to be rather straightforward. There is a strong correlation between colour and 
the mass indicator ($\sigma_e$), analogous to the $V-K$, $J-K$ and 
$U-V$ relations of Bower, Lucey \& Ellis (1992), and the $V$ --[3.6] vs. $\sigma$ relation of Paper XIX.  The main
difference is that the $[3.6]-[4.5]$ colour becomes {\it bluer} for increasing galaxy
mass/luminosity. Before we can interpret this fact, we
first need to look at the behavior of the [3.6] - [4.5] colour in some more detail.

\onecolumn

\twocolumn

\subsection{Clues on interpreting the [3.6]--[4.5] colour.}
\label{young}

A rough comparison with the stellar population models shows that none of these sets of models is very suitable
to explain our observations. For stellar populations above 2 Gyr, the models of
 Charlot \& Bruzual (2007) are probabaly not accurate enough, since they do not reproduce the sign of the
observed colour-metallicity relation. The models of Marigo et al. (2008) are also
not very adequate, since they also do not show much blueing as a function of
metallicity at high metallicities. At least, their predicted colours are more or
less in agreement with the observations: [3.6] - [4.5] $\sim$ --0.1 for old
stellar populations. The models below 2 Gyr are obviously much more complicated,
making it hard to believe any precise predictions here. What we can say,
probably, is that the $[3.6] - [4.5]$  colour fluctuates considerably with age
and metallicity, that very blue colours  can in principle be explained by
adding young AGB stars.

\begin{figure}
\begin{center}
\includegraphics{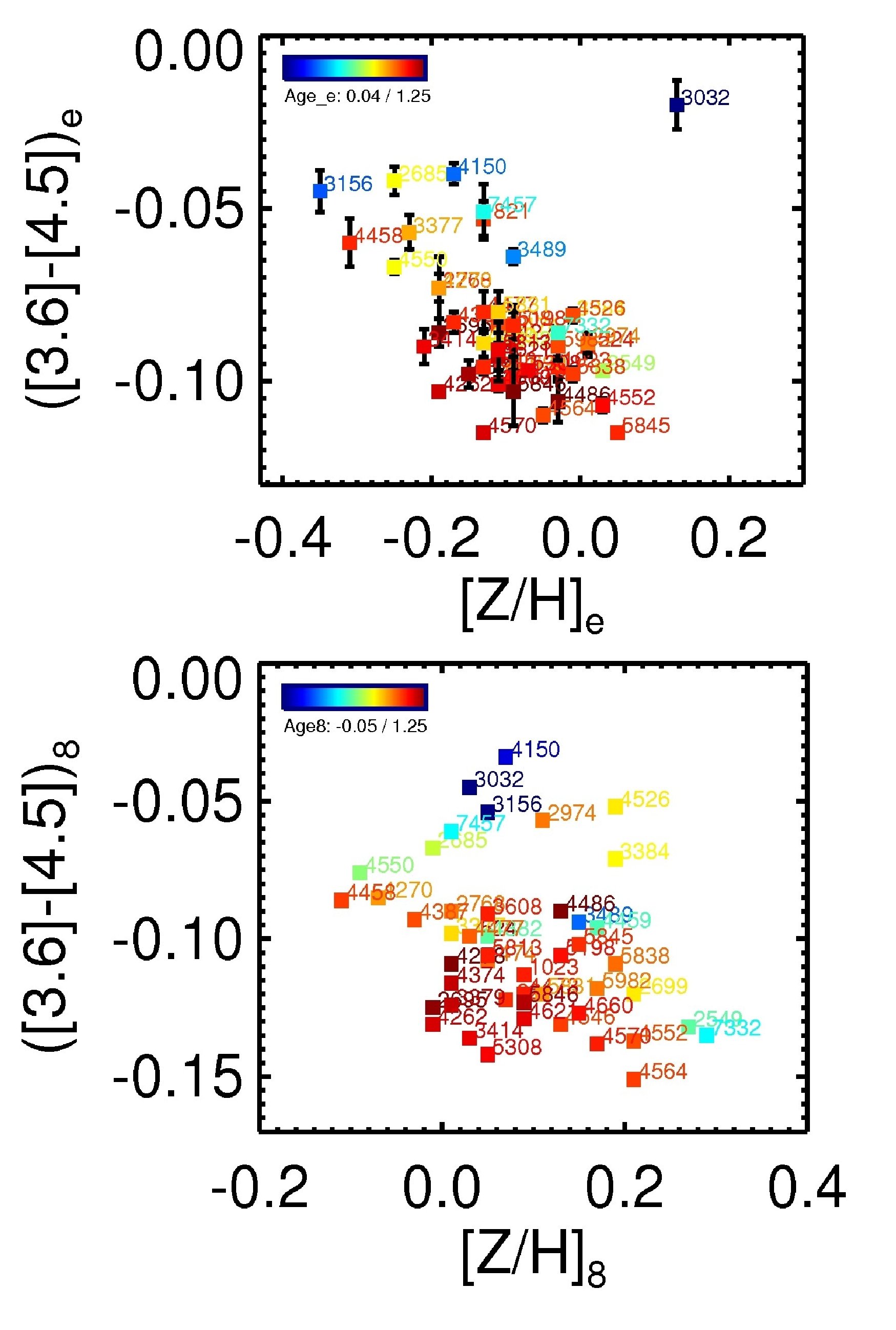}
\caption{Relation between SSP-metallicity [Z/H] (Paper XVII) and the [3.6] -
[4.5] colour inside  r$_e$ (Fig.\ref{colmet}a) and r$_e$/8 (Fig.\ref{colmet}b).
Colouring of the points as in Fig.\ref{mass_met}. } 
\label{colmet}	 
\end{center}
\end{figure}

Next, we make an empirical comparison between the [3.6] - [4.5] colour and the
metallicity, from Paper XVII, obtained using SSP fitting to the SAURON spectra
(Fig.~\ref{colmet}).
It is clear that the scatter  for the colours inside r$_e$/8 is larger than inside r$_e$. The reason
for this is that in the center one often sees a combination of old and young
stellar populations at the same place. When one then analyzes the combined
stellar populations as stellar population with a single age and metallicity, one
can obtain metallicities that are too large (e.g. Allard et al. 2006). An
extreme example here is NGC~3032. Concentrating on Fig.~\ref{colmet}a, one sees
that there is a good correlation between metallicity [Z/H]  and the [3.6] -
[4.5] colour, indicating that for this sample the [3.6] - [4.5] colour is a
good  metallicity-indicator. Younger galaxies, are generally found
above the mean relation, indicating that younger stellar populations cause the
[3.6] - [4.5] colour to redden.

\begin{figure}
\centering
\includegraphics[width=9truecm,clip]{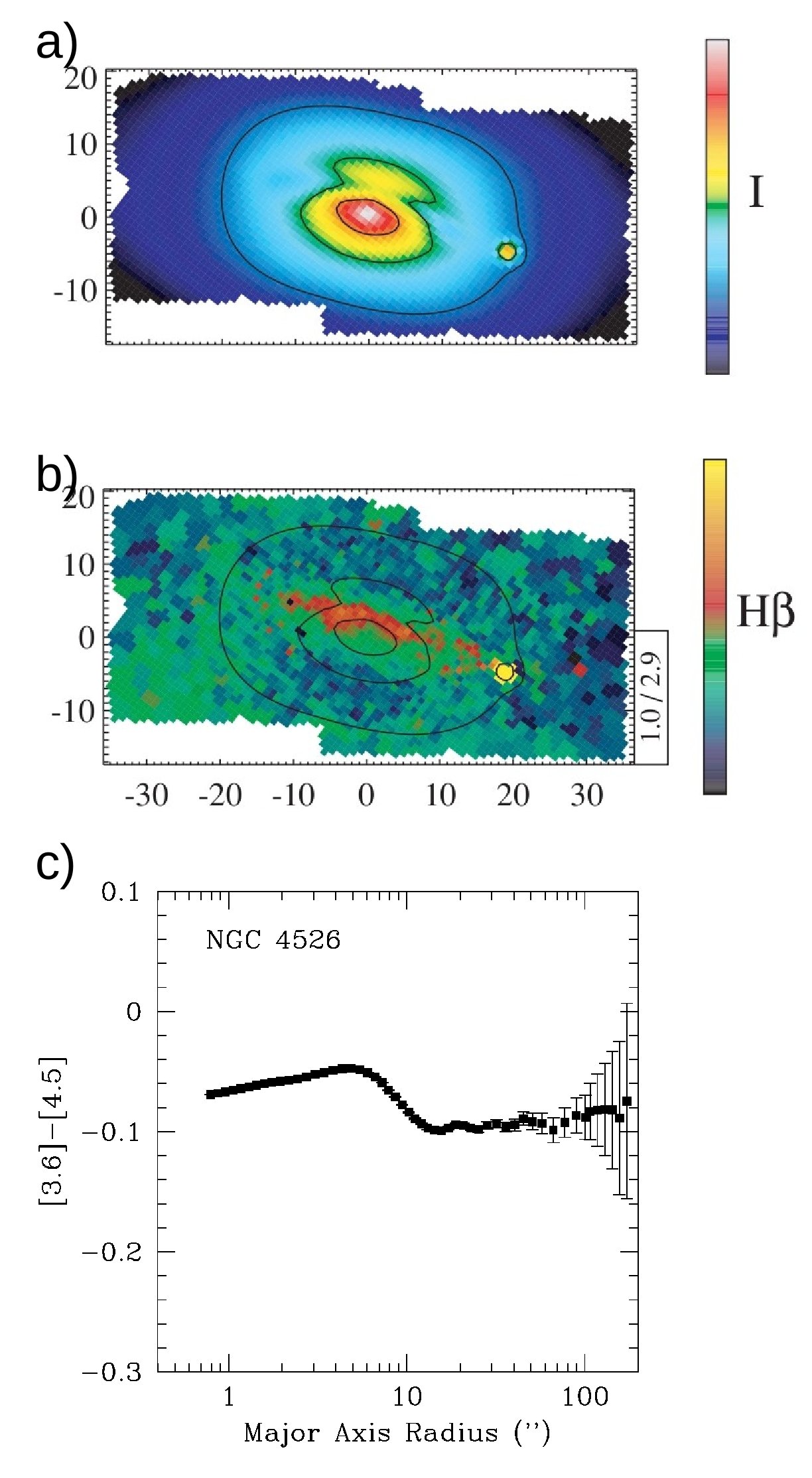}
\caption{SAURON images in $V$-band continuum (a)) and H$\beta$ absorption 
in \AA  (b)) for NGC~4526. The yellow spot is a foreground star and does not reflect truly strong H$\beta$ in the galaxy. In the bottom
panel our [3.6] - [4.5] profile is shown. The central redder part of the profile corresponds exactly
to the inner disk seen in the upper 2 panels.} 
\label{n4526}
\end{figure}

An outlier in  Fig.~\ref{mass_met8} is NGC~4526. It is
instructive to study its colour profile, to understand why this galaxy is an
outlier. In Fig.~\ref{n4526}c one can see that the galaxy has a
regular colour profile, the colour becoming redder slowly going outward, 
starting at about 12$''$. Inside this radius the colour is significantly redder. In the upper 2 panels of the same figure one can see why this is the
case. The galaxy has a disk in the inner 12$''$, which can be seen in 
extinction in the continuum image. The disk also contains younger stellar
populations, which can be seen from its high values in the H$\beta$ 
absorption image. Note that this image has been corrected for emission: the
inner disk also contains strong emission lines (see Paper VI). As a result of the young stars, 
the CO-absorption band in the 4.5 $\mu$m filter becomes less deep,
causing the [3.6] - [4.5] colour to redden.
The same effect is seen in many other galaxies, where the presence of young stars in
the line strength images of paper VI, or the presence of H$\beta$ in emission
in Paper V, is often accompanied by a redder [3.6] - [4.5] colour. Examples are NGC
3032, NGC 3384, NGC 4459 and NGC 5838. Since the young populations generally are concentrated
only in the central areas, they affect the colours within $r_e/8$ much more than
the one within $r_e$. 

\subsection{Central point sources}
\label{sec:agn}

Inspecting Fig.\ref{mass_met8} we see a number of outliers
above the colour-$\sigma$ relation, that have [3.6] - [4.5] colours  redder than
expected for their $\sigma$. Although this could be due to contributions from
young stellar populations, it could also be that non-thermal light from AGNs is
causing this effect. The prime example here in M87 (NGC 4486), for which there
is no independent evidence for young stellar populations. Other options are NGC
2768 and 2974. NGC~4486 especially is remarkable.  In Fig.~\ref{prf2} one sees
that this galaxy has a very red central pointsource, while for the rest of the
profile the colour is almost constant. Its colour profile has been made while
masking out the jet, but not the center. In the optical this galaxy shows a blue central point source (e.g. Biretta et al. 1991). AGNs
generally show a synchrotron spectrum, appearing bluer than the stellar light in
the optical and  redder in the infrared (Polletta et al. 2006). Tang et al.
(2009) studied Spitzer IRAC colours of a sample of 36 local elliptical galaxies.
They detect non-stellar infrared emission in 9 of them. All of these show
AGN-like optical spectra, and are classified as 8 Liners and 1 transition object
using the classification of Ho et al. (1997). Not all AGNs observed, however,
show this infrared excess. Tang et al. conclude that this infrared excess
appears only and universally in galaxies with a relatively luminous central
AGN. 

\begin{figure}
\centering
\includegraphics[width=9truecm,clip]{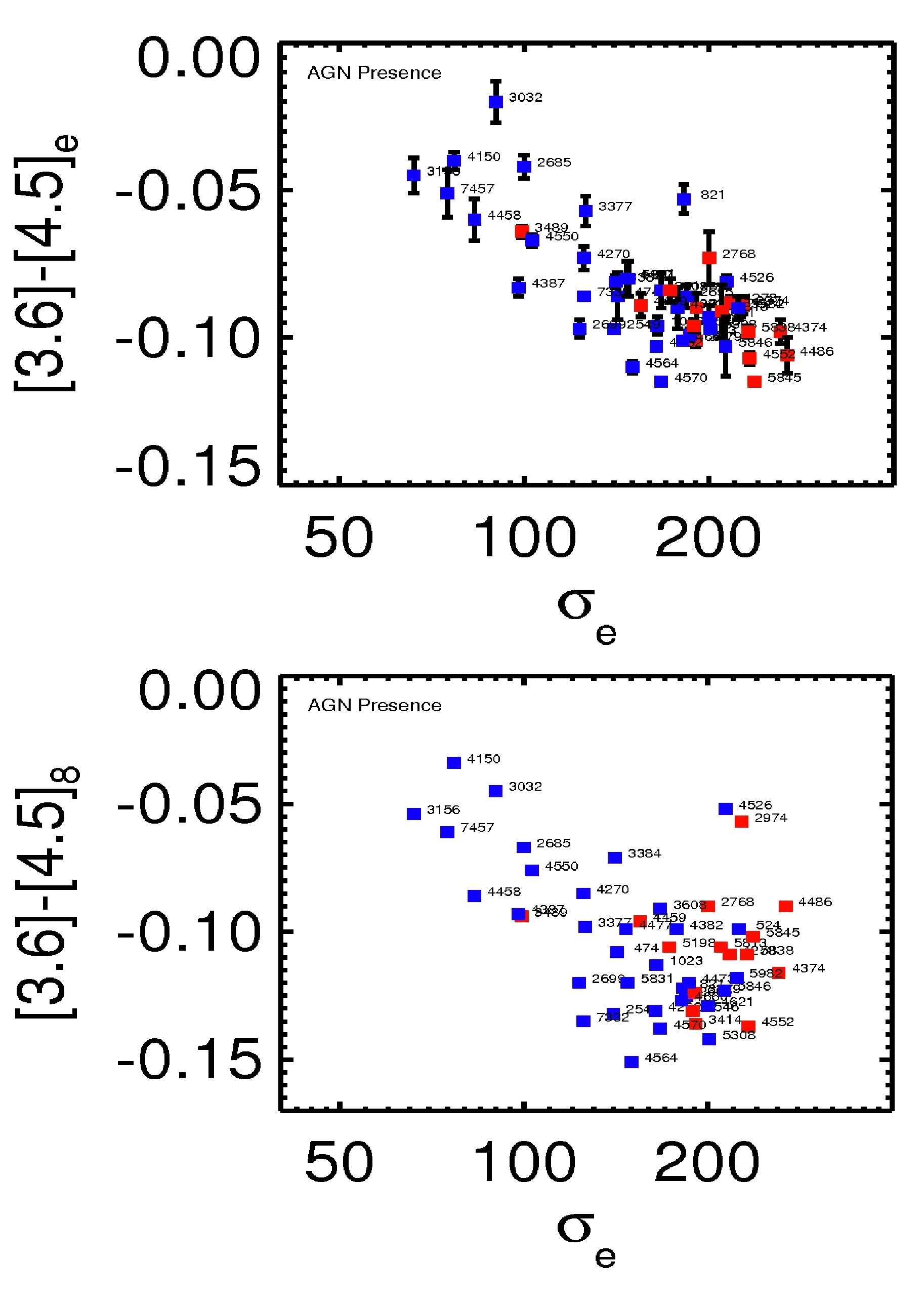}
\caption{Colour - $\sigma$  (in km/s) relation, as in Fig.~\ref{mass_met}. Here galaxies containing compact central radio sources, indicative of AGNs,
detected by the  VLA FIRST survey (Becker, White \& Helfand 1995), are plotted in red.} 
\label{agnfig}
\end{figure}

Here we try to reproduce their result. We have selected the AGNs in our sample by selecting all 15 galaxies that have a compact radio
source, as detected by the  VLA FIRST survey (Becker, White \& Helfand 1995, Paper XVI). These galaxies have been represented in red in
Fig.~\ref{agnfig}. One sees that the galaxies that contain an AGN preferentially lie above the colour - $\sigma$ relation. This effect is
stronger in Fig.~\ref{agnfig}b, where the colour inside r$_e$/8 is given, which is obviously more affected by the AGN than the colour inside
r$_e$. We are seeing that the integrated colour is redder than for galaxies without AGNs. This could be due to the non-thermal light of the
AGN (e.g. in NGC~4486) or the hot dust around it, but also due to younger stars near the center, which dilute the 4.5 $\mu$m feature. The
red light clearly is centrally concentrated, which agrees with the fact that Sarzi et al. (2010, Paper XVI) conclude that a central AGN
source could be responsible for the ionization of the gas in the very central regions (2-3$''$). The galaxies with an AGN are mostly the
largest galaxies - they tend to be dominated by old stellar populations. The fact that in optical line indices the effects of the AGN are
generally not seen might indicate that we are seeing here the hot dust (or non-thermal emission) around the AGN.

In a paper studying all Spitzer-IRAC colours of the SAURON-Sa sample Van der Wolk et al. (in preparation) discuss the behaviour of AGN in
this complementary sample in more detail, using all four IRAC colours.

\subsection{Scatter in the [3.6] -- [4.5] -- $\sigma$ relation}
\label{sec:scatter}

\begin{figure}
\begin{center}
\includegraphics[width=\textwidth]{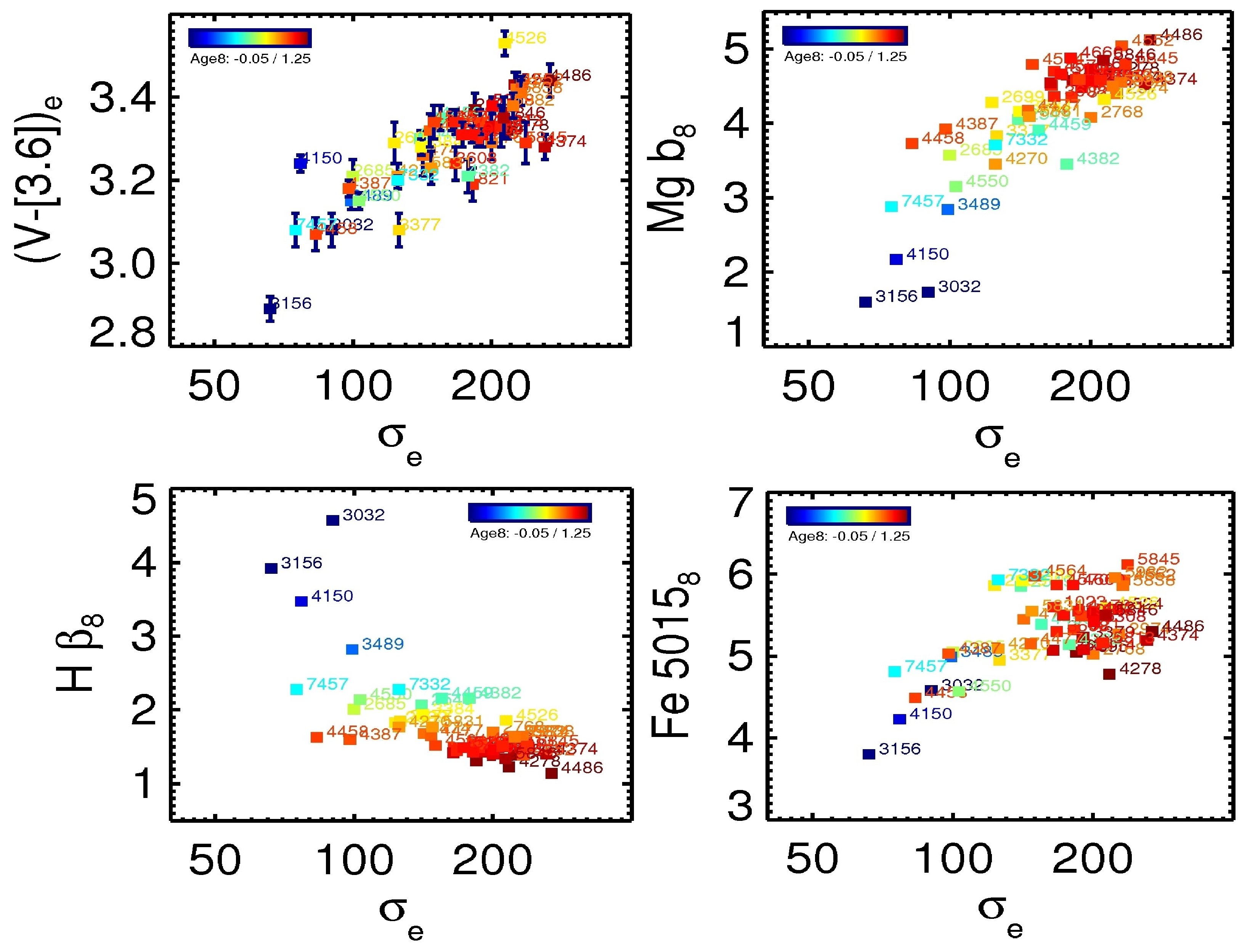}
\caption{Various colour/line strength vs. velocity dispersion plots. In Fig.~\ref{mass_ls}a (top-left) 
the $V$-[3.6] colour in mag~arcsec$^{-2}$  within r$_e$ is plotted (from Paper XIX).  
All line strengths (in Fig.~\ref{mass_ls}b,c and d)  are equivalent widths in \AA\ and
have been calculated within apertures of $r_e/8$. The points are coloured according to age within $r_e/8$.
} 
\label{mass_ls}	 
\end{center}
\end{figure}

If the scatter in the colour-$\sigma$ relation can be explained by young stellar
populations on top of a much older underlying stellar population, we would
expect the outliers of the optical  line strength - $\sigma$ relations of Paper
VI to be the same as the outliers of the colour - $\sigma$ relation here. In
Paper VI a tight Mg~{\it b} - $\sigma$ relation is shown for the galaxies in the
SAURON sample. The scatter is somewhat larger than e.g. in the Coma cluster
(Smith et al. 2009) due to the fact that there are several galaxies in the
SAURON sample with young central stellar populations. In Fig.~\ref{mass_ls}
three of the relations of Paper VI are shown, together with the $V$-[3.6] --
$\sigma$ relation (Fig. 7 of Paper XIX, but expanded).  One sees that the
galaxies show a tight Mg~{\it b} -- $\sigma$ relation, with some galaxies
falling below the average relation. For example, the Mg~{\it b} values of  
NGC~4150 and NGC~3032 are lower than those of NGC~7457, a galaxy with a similar
velocity dispersion. The reason is that a considerable young stellar population
is present in both NGC~4150 and NGC~3032 (Paper VI, Paper XVII). The galaxies
that lie below the Mg~{\it b} -- $\sigma$ relation generally lie above the
H$\beta$ -- $\sigma$ relation. For Fe 5015 the situation is much less clear. The
scatter in the $V$-[3.6] vs. $\sigma$ relation is due to extinction (making the
colour redder) and young stellar populations (making it bluer). The outliers in
this diagram are not necessarily the outliers of the others.

\begin{figure}
\centering
\includegraphics[height=0.9\textheight,clip]{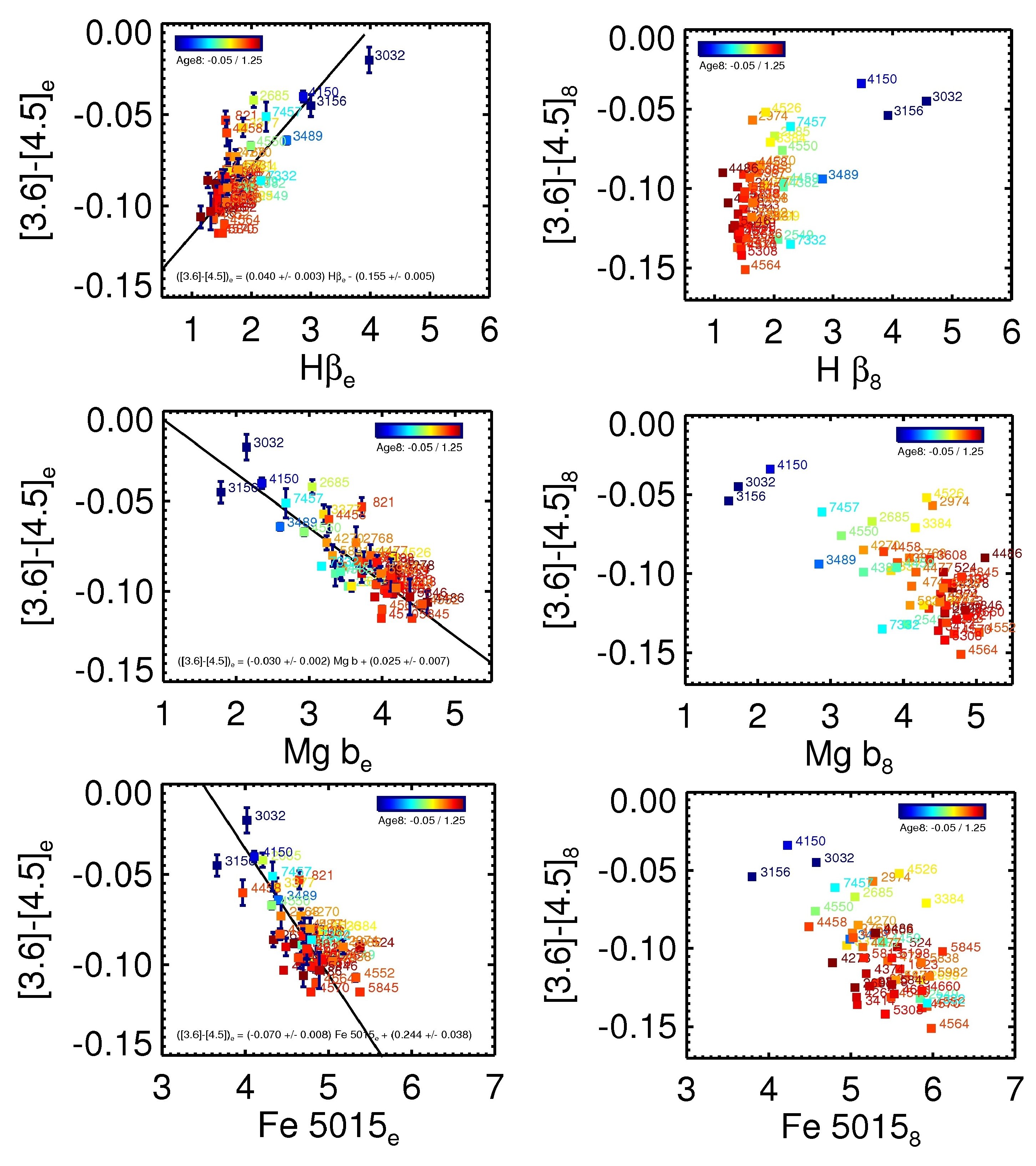}
\caption{$[3.6]-[4.5]$ colours  in mag~arcsec$^{-2}$ measured inside $r=r_e$ and $r_e/8$ versus
line strength indices  (equivalent widths, in \AA\ )  (from Paper VI) measured within the same circular
apertures.} 
\label{indices}
\end{figure}

In general, the [3.6] - [4.5] vs. $\sigma$ diagram agrees with the other line strength -- $\sigma$ diagrams. 
Young stellar populations make the [3.6] - [4.5] color redder, H$\beta$ stronger, 
and Mg~{\it b} and Fe 5015  weaker. There is no direct correlation between the 
residuals, since every relation has a different curvature.
Some galaxies, such as NGC~4486, only deviate from
the colour - $\sigma$ relation. As mentioned before, this is
probably due to the presence of non-thermal emission from AGNs. The agreement 
is quite remarkable. For example, galaxies that lie below the H$\beta$ -- $\sigma$ 
relation also lie below the [3.6] - [4.5] -- $\sigma$ diagram. All this causes the 
colour -- line strength relations (Fig.~\ref{indices}) to be rather straight.
It looks as if just metallicity and age are fully responsible for the positions of the 
galaxies in these colour/line strength -- $\sigma$ relations (with the exception of NGC 4486, 
and dust extinction in V - [3.6] vs. $\sigma$). Although one would a priori expect 
AGB light to be of considerable influence to the [3.6] - [4.5] colour, and not to the 
optical line indices, its effects are not found in this study.

We proceeded to try to find out whether the scatter in the $[3.6] - [4.5]$  vs.
$\sigma$ relation could be due to other parameters. To do this,  we performed a principle
component analysis on many galaxy parameters. Although this analysis shows that many parameters correlate with
 $[3.6]-[4.5]_e$ , namely $\sigma_e$, 
Age and $\alpha$/Fe within r$_e$ (from Paper XVII), the Spitzer [3.6] luminosity M$_{3.6}$ and $V - [3.6]_e$ (both from
paper XIX), dynamical mass and M/L (both from Paper XIV), there is not much power in the second, third and higher
eigenvectors (46\% in EV1, 15\% in EV2, 11\% in EV3, 7\% in EV4 etc.). Furthermore, these eigenvectors are hard to relate to
the physical parameters.  From EV2 there is a hint that the 
residuals of the $[3.6] - [4.5]$  vs. $\sigma$ relation might correlate with r$_e$ (from paper XIX), 
boxy/diskiness $a4/a$ (from
Paper IX) and M$_{HII}$, the amount of ionized gas present in the galaxy (from Paper V). This is 
understandable if this effect is caused by the presence of central young disks: galaxies with such disks 
show larger $a4/a$, have more ionized gas, and are generally smaller than other galaxies. +0.1543=
No correlations are 
found with either  morphological type, ellipticity, $\lambda_R$, a parameter measuring
the angular momentum in a galaxy, (both from Paper IX), metallicity 
within r$_e$ (from paper XVII), the FUV luminosity  (Bureau et al., Paper XVIII),
and the galaxy environment (cluster or field, following the definition of Paper
II).

\onecolumn

\twocolumn


\subsection{Colour - mass and colour - magnitude relations}

Since we have calculated the total mass of the galaxies (in Paper XIV) we also
show the relations  $[3.6]-[4.5]$ colour and mass and with 3.6 $\mu$m luminosity
in Fig.~\ref{colmb}.  Although these quantities are dependent on distance, the
colour-mass relation might have a more physical meaning. The fact that these
relations show considerably more curvature than the colour - $\sigma$ relation,
might give some clues about their origin. The curvature is the same as seen in
the colour - metallicity relation {Fig.~\ref{colmet}).  The difference is due to
a curvature in the relation between log(mass) and log($\sigma$), together with
the fact that some galaxies, such as NGC 4564, NGC 4570 and NGC 5845, have a
high $\sigma_e$ for their mass or luminosity. These objects remain old, and
have  [3.6] - [4.5] colour, but are considerably displaced along the mass-axis.
The reason for the curvature at the low mass end is the curvature in the mass vs.
$\sigma$ relation, or the Faber-Jackson relation (see e.g. Paper XIX). At the same $\sigma$
compact galaxies such as NGC~5845 are much less massive than others. 

\begin{figure}
\begin{center}
\includegraphics{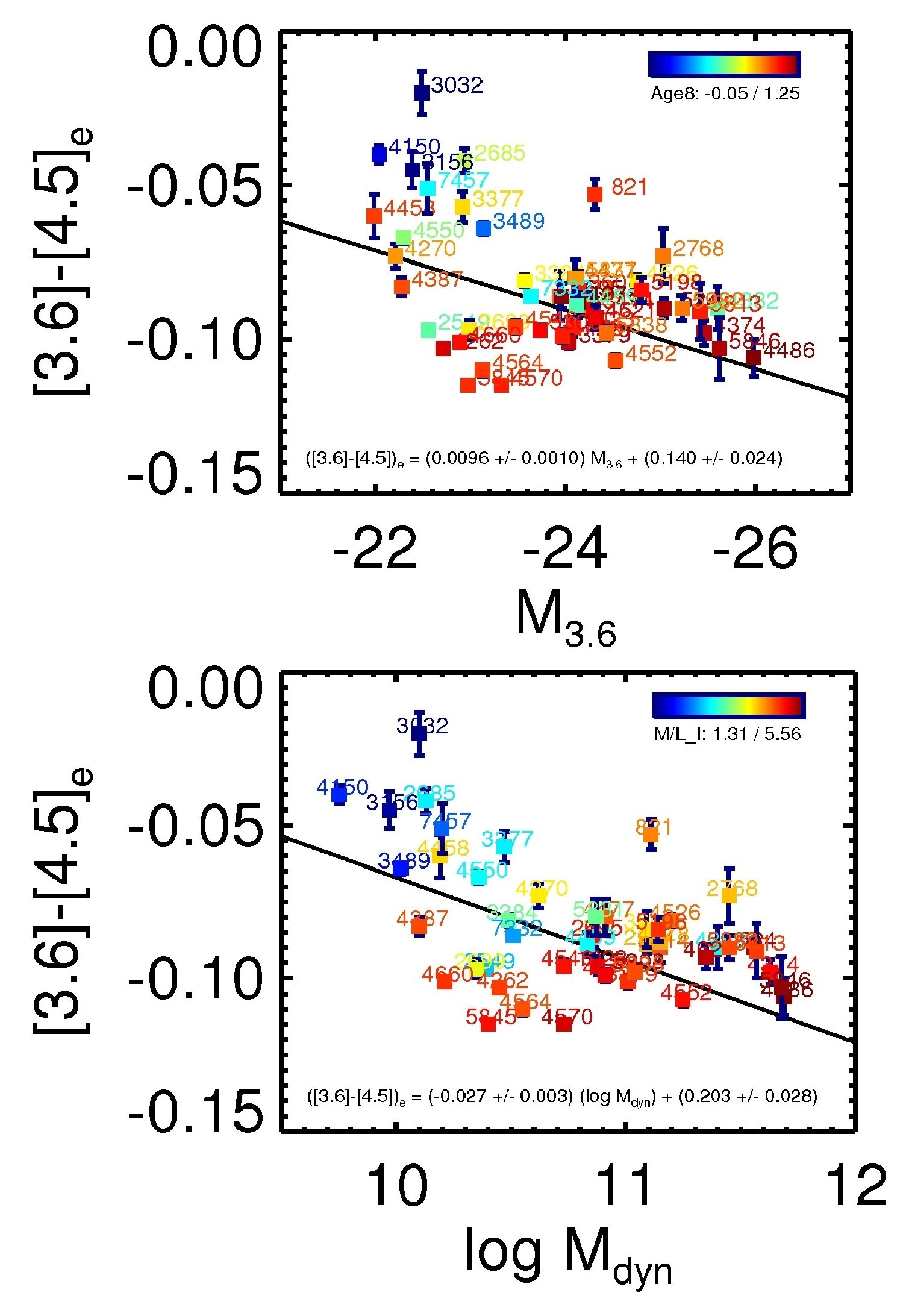}
\caption{Relation between a) absolute 3.6 $\mu$m magnitude from paper XIX and colour and b) mass (from paper XIV,
in solar mass) and colour. The
point are coloured according to their age within r$_e$/8. In the bottom figure the points have been coloured according to the M/L inside r$_e$ (from Paper XIV).} 
\label{colmb}	 
\end{center}
\end{figure}

\subsection{Colour-colour and colour-line strength relations}
\label{sec:linerelations}

We now present the $[3.6]-[4.5]$ $\mu$m colour as a function of the three line
strength indices published in Paper VI (Fig.~\ref{indices}). Values inside $r_e$
are presented on the left. Strong correlations are found for all indices,
especially inside $r_e$. This is expected, since the indices themselves also
correlate well with $\sigma$ (Paper VI). Galaxies deviating from the trends are
those with AGNs, and those with a considerable amount of younger stellar
populations: adding young stellar populations to a galaxy with old stellar
populations does not always make it move along the average colour - index relation.

Here we try to interpret the data in a very rough way using the models of Marigo
et al. (2008). In Fig.~\ref{mgb_colour} we have plotted these models on top of
our data. Models are plotted for metallicities Z=0.008, 0.02 and 0.03
(triangles, circles and filled squares). The age of the models is indicated by
the colours of the points. The first thing to note is that for old stellar
populations the model colours become bluer and Mg~{\it b} stronger with increasing
metallicity. This is in agreement with the data. However, for constant
metallicity, and models above about 2 Gyr, stellar populations have the same
colour, while Mg~{\it b} is decreasing when age decreases. This does not seem to
be in agreement with the observations. In e.g. the central regions of NGC~4526,
where we find clear evidence for younger stellar populations, the [3.6] - [4.5]
$\mu$ colour becomes considerably redder. 

We find that galaxies within $r_e/8$ both have a bluer colour and a stronger
Mg~{\it b} index than within $r_e$.  This can be explained easily by a gradient
in metallicity. It remains, however, to be seen to what extent this diagram can
be used to separate age and metallicity. To do this, a prerequisite is that much
more accurate stellar population models are available for the [3.6] - [4.5]
colour. 

Plotting [3.6] - [4.5] against $V$ - [3.6] in Fig.~\ref{colcol} we see that there is a reasonably tight
correlation between  both colours. As mentioned in Subsection \ref{sec:scatter}
the scatter is caused by dust extinction in the V-band, and multiple stellar
populations. 

\begin{figure}
\centering
\includegraphics[width=0.49\textwidth,clip]{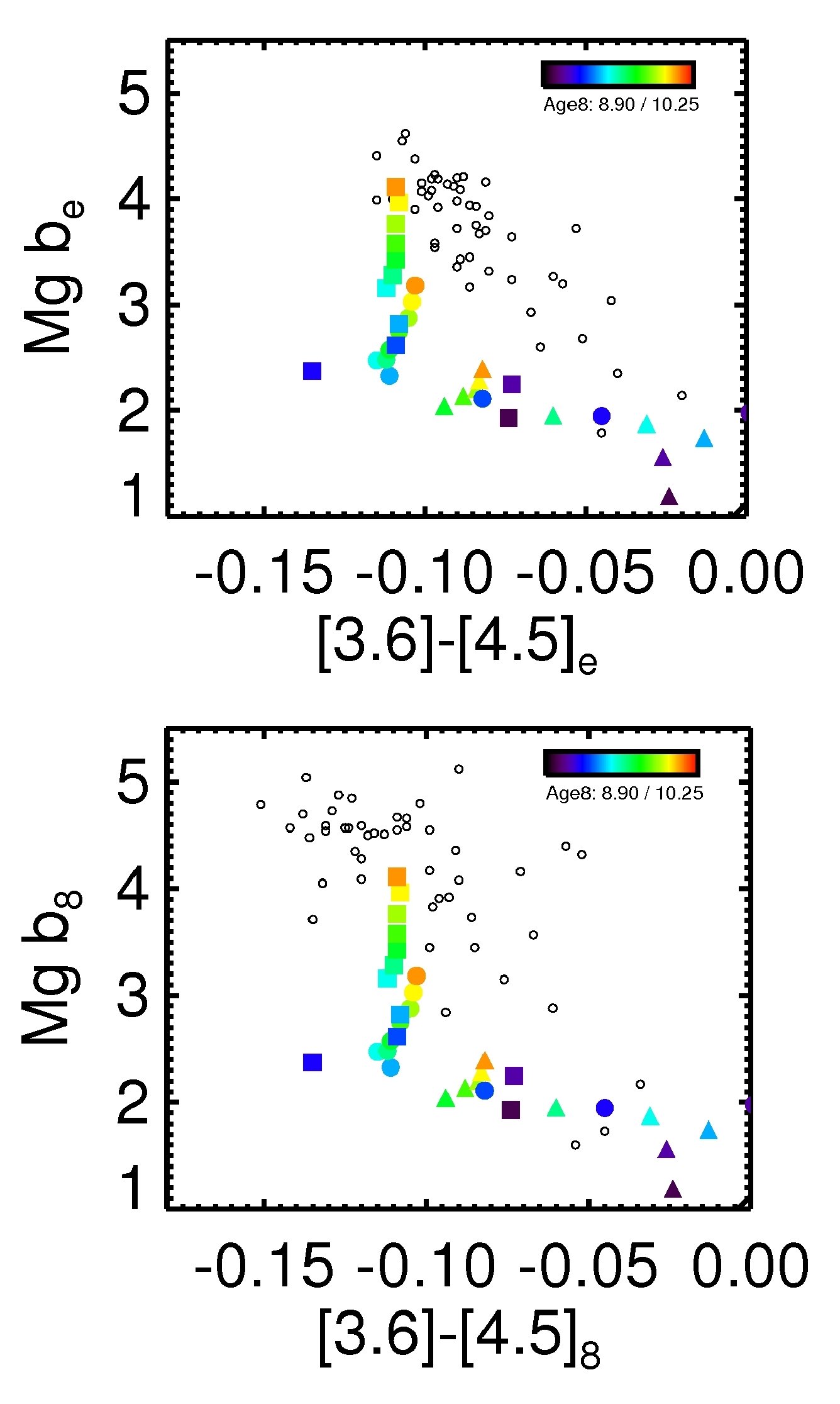}
\caption{$[3.6]-[4.5]$ colour - Mg~{\it b} diagrams. The black open circles indicate
the galaxy data presented in Section 4 using apertures of radius r$_e$ (top)
and r$_e$/8 (bottom). Coloured points indicate SSP models, composed of
$[3.6]-[4.5]$ colours from Marigo et al. (2008) and Mg~{\it b} line indices from
Bruzual \& Charlot (2003). Models are given for solar metallicity Z=0.008 (filled triangles), Z=0.02 (filled circles) ans Z=0.03 (filled
squares). The models
are coloured according to their age. }
\label{mgb_colour}
\end{figure}

\begin{figure}
\begin{center}
\includegraphics{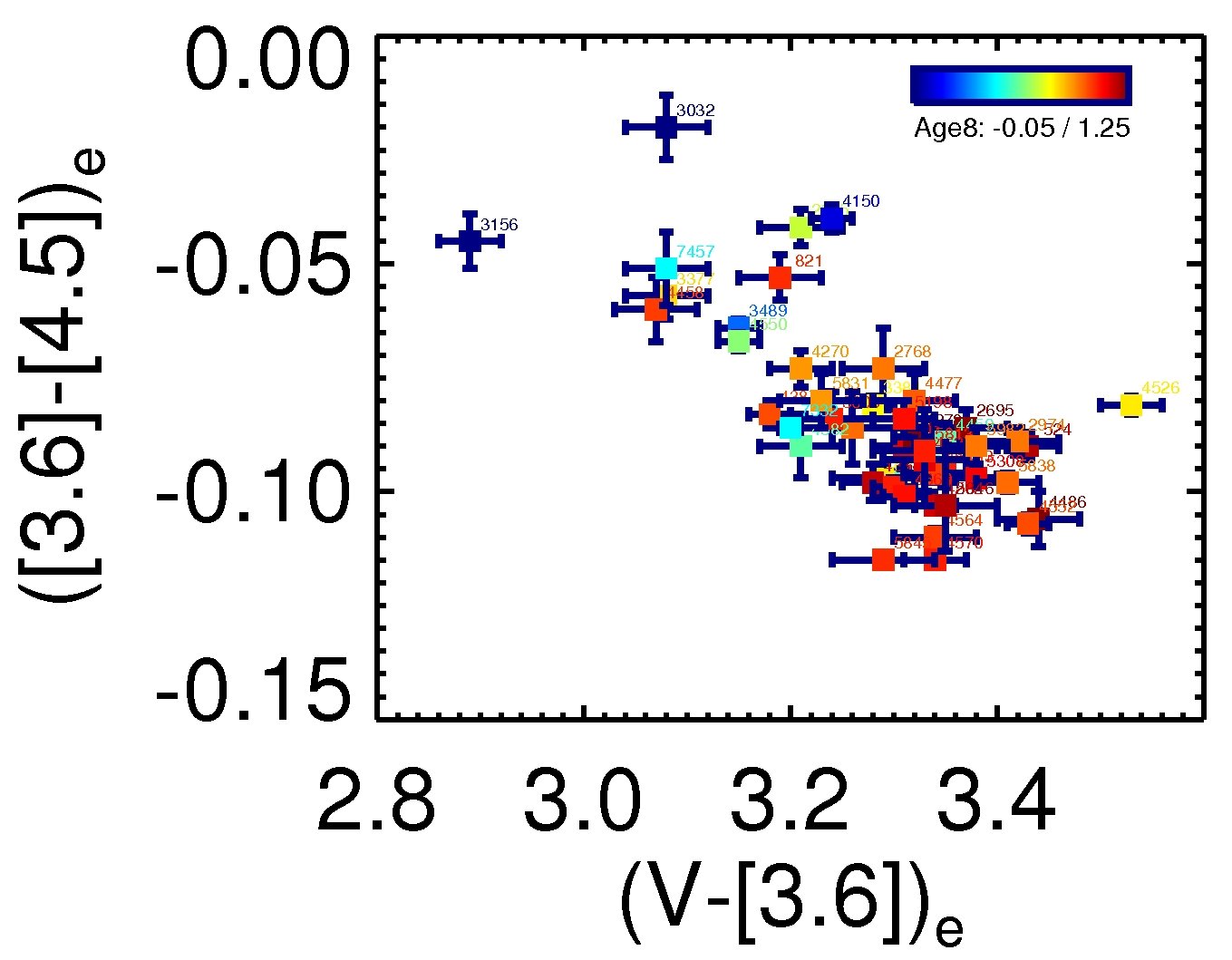}
\caption{[3.6] - [4.5] colour vs. $V$ - [3.6] (from Paper XIX, both in mag~arcsec$^{-2}$) within r$_e$.}
\label{colcol}	 
\end{center}
\end{figure}

\section{Local colour relations}

\subsection{What are we measuring?}

Early-type galaxies generally show small radial colour gradients in the optical
and near-infrared (e.g. Peletier et al. 1990). These gradients reflect stellar
population gradients, but can also be caused by dust extinction. In general,
stellar population gradients can be caused by gradients in metallicity, and by
radial age differences. In giant ellipticals, it is generally thought that
galaxies have a weak metallicity gradient ($\Delta(Z/H)/\Delta(\log~r)$ $\sim$
-0.4), with average metallicity decreasing slowly going outward. On top of that,
often indications of younger stellar populations are found in the inner regions
(see e.g. Fig. \ref{n4526} and Paper XVII). These younger stellar populations
are then generally accompanied by dust extinction, spiral arms, and/or star
forming regions, although in the MIR dust extinction does not play a role (Martin \& Whittet 1990).
The presence and amount of young stellar populations depends on the
environment.  S\'anchez-Bl\'azquez et al. (2006) find age gradients for their
sample of galaxies in low density environments, but their results for galaxies
in a high density environment (Coma) point at age gradients consistent with
zero. However, Rawle et al. (2010) warn that, even though they find on average
a zero age gradient, 40\% of their sample is inconsistent with the absence of an
age gradient. This agrees well with the SAURON sample, in which several
galaxies have central regions that are much younger than the outer parts (Paper
XVII).  With these results in mind, we define for each galaxy a region in which
we find evidence for younger stellar populations. Outside this region, we fit a
linear relation between [3.6] - [4.5] colour and logarithmic radius (see Section
3.3). The slope of this relation is the colour gradient, which is (by design) likely to be
caused by a change in  metallicity. Note that all colour gradients are
positive. 

We now look at the relation between the $[3.6]-[4.5]$ colour gradients and
gradients in the three indices that can be measured by SAURON: Mg~{\it b},
H$\beta$ and Fe 5015 (Paper VI). Here we use a simple, robust definition for
gradients, both for colours and indices: we compare differences between the
colours/indices in the smaller aperture of $r=r_e/8$ and the one of $r=r_e$   
We find that the $[3.6]-[4.5]$ colour difference correlates reasonably well with
the Mg~{\it b} difference, that the scatter increases for Fe 5015, and that
there is no correlation with H$\beta$. All this is consistent with the
integrated colour - index relations (Section 4.7).

Since the colour gradients have been constructed to measure metallicity
gradients, we now compare them with linear fits to the radial metallicity
profiles [Z/H] from Paper XVII. These SSP-metallicities have been obtained by
fitting SSP-models to the azimuthally averaged line indices of the SAURON
galaxies. In Figure~\ref{metallicity_grad} we find a good correlation. The
points, coloured by the age inside r$_e$/8, show that galaxies with large colour
gradients also have large metallicity gradients. On the other hand, there are
some objects with large metallicity gradients for which the colour gradient is
much smaller than expected from the other galaxies. The colouring of the points
shows that this mostly happens for galaxies with younger central ages, so this
might indicate that the colour gradients in the galaxies with younger centers
are still affected by those younger stellar populations, even though care has
been taken to reject the regions affected.  Circumstantial evidence for this
comes from the 8 $\mu$m maps (Shapiro et al. 2010, Paper XV), where
stellar-corrected 8 micron emission is detected up to large radii in all
outliers in Fig.~\ref{metallicity_grad} . The presence of this 8 $\mu$m flux
comes from warm dust, which could well be accompanied by nearby star formation.
In this case the metallicity gradients is probably diluted by the presence of
intermediate-age stellar populations ($10^8$ - $10^9$ year), which dominate the
mid-infrared region around 4 micron. Very detailed statements about this cannot
be made yet, since the range in [3.6] - [4.5] colour covered by the galaxies is
very limited, so e.g. a morphological decomposition of galaxies into an old and
an intermediate age component is not possible with the current data.

\begin{figure}
\centering
\includegraphics[width=0.48\textwidth]{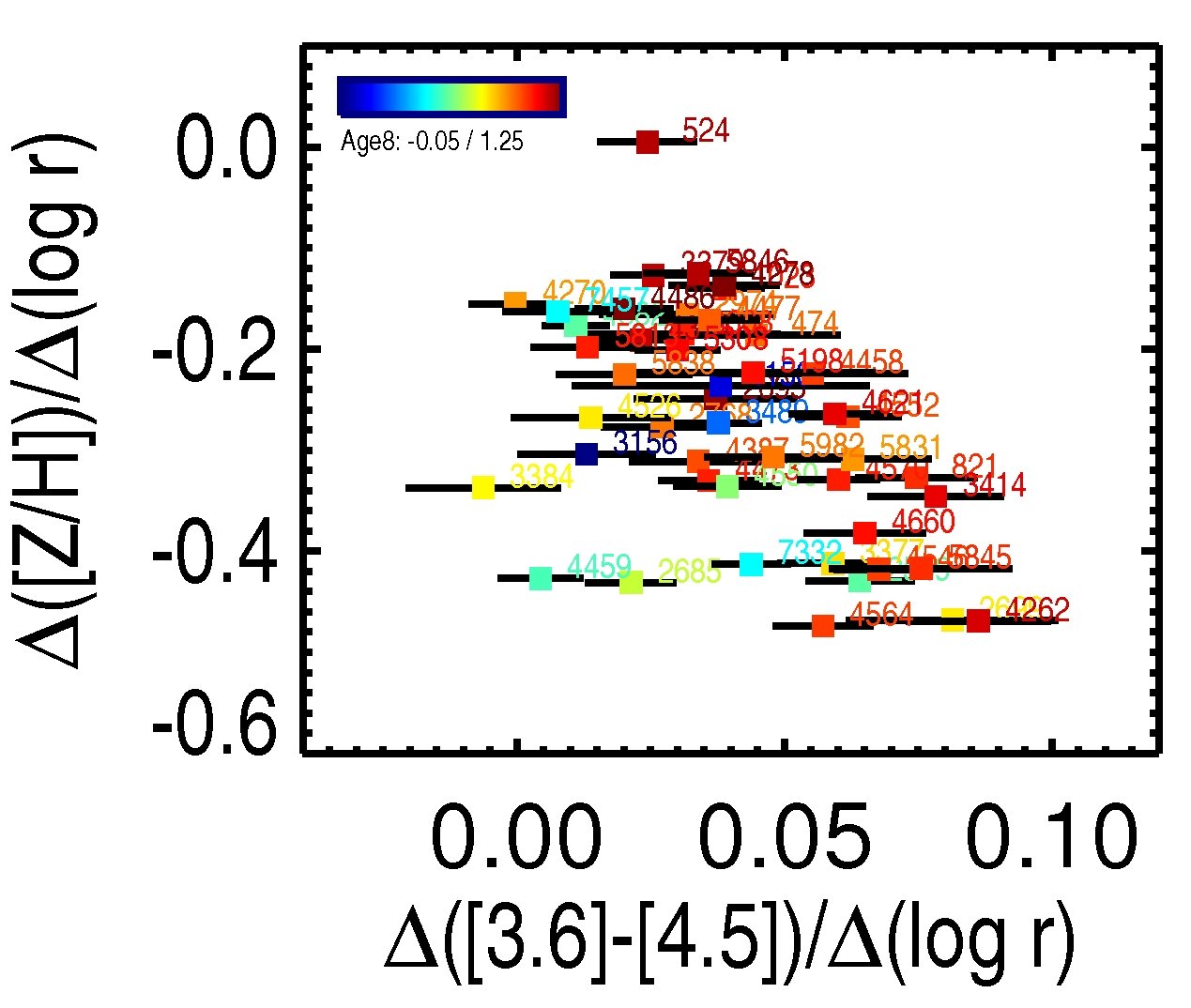}
\caption{Metallicity gradients from Paper XVII, plotted against the [3.6]-[4.5] colour gradients. 
The points have been coloured according to the SSP-age inside r$_e$/8.} 
\label{metallicity_grad}	 
\end{figure}

\subsection{Dependence of colour gradients on galaxy mass}

In a recent paper (den Brok et al. 2011) we discuss stellar population gradients
in early-type galaxies covering a large range in mass in the Coma cluster. We
find that colour gradients in general correlate with galaxy luminosity or mass,
in the sense that high-mass galaxies have stronger gradients than dwarfs. A new
result in that paper is that optical colour gradients in the main body of the
galaxies do not become positive for very faint galaxies, as has been claimed in
the literature (e.g. Vader et al. 1988). The reason why generally positive
gradients are measured when using the whole available  range in radius, is due
to the central component, which is very important for the faintest galaxies.
This central component, which in dwarfs is called a nuclear cluster, is often
bluer  than the main galaxy, making the colour gradients change sign. In den
Brok et al. a detailed comparison is also given with the literature on stellar
population gradients. In the current paper we only discuss bright galaxies, a
subset of the range in mass covered in den Brok et al. Our sample, however, has
been studied  in much more detail, and for that reason knowing their metallicity
gradients might more easily lead to a better understanding of the formation
processes of early-type galaxies.

\begin{figure}
\begin{center}   
\includegraphics[width=0.39\textwidth]{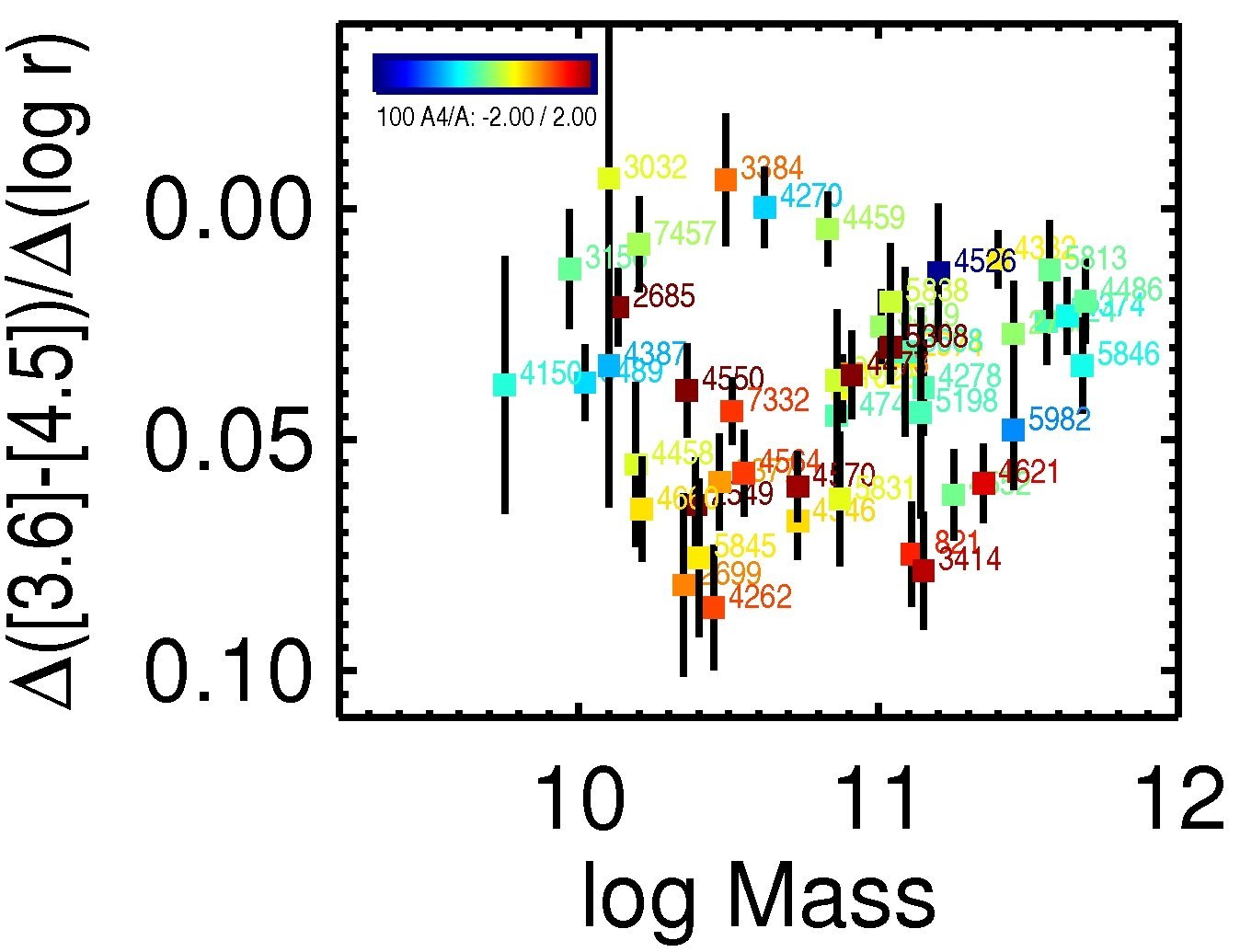}
\includegraphics[width=0.39\textwidth]{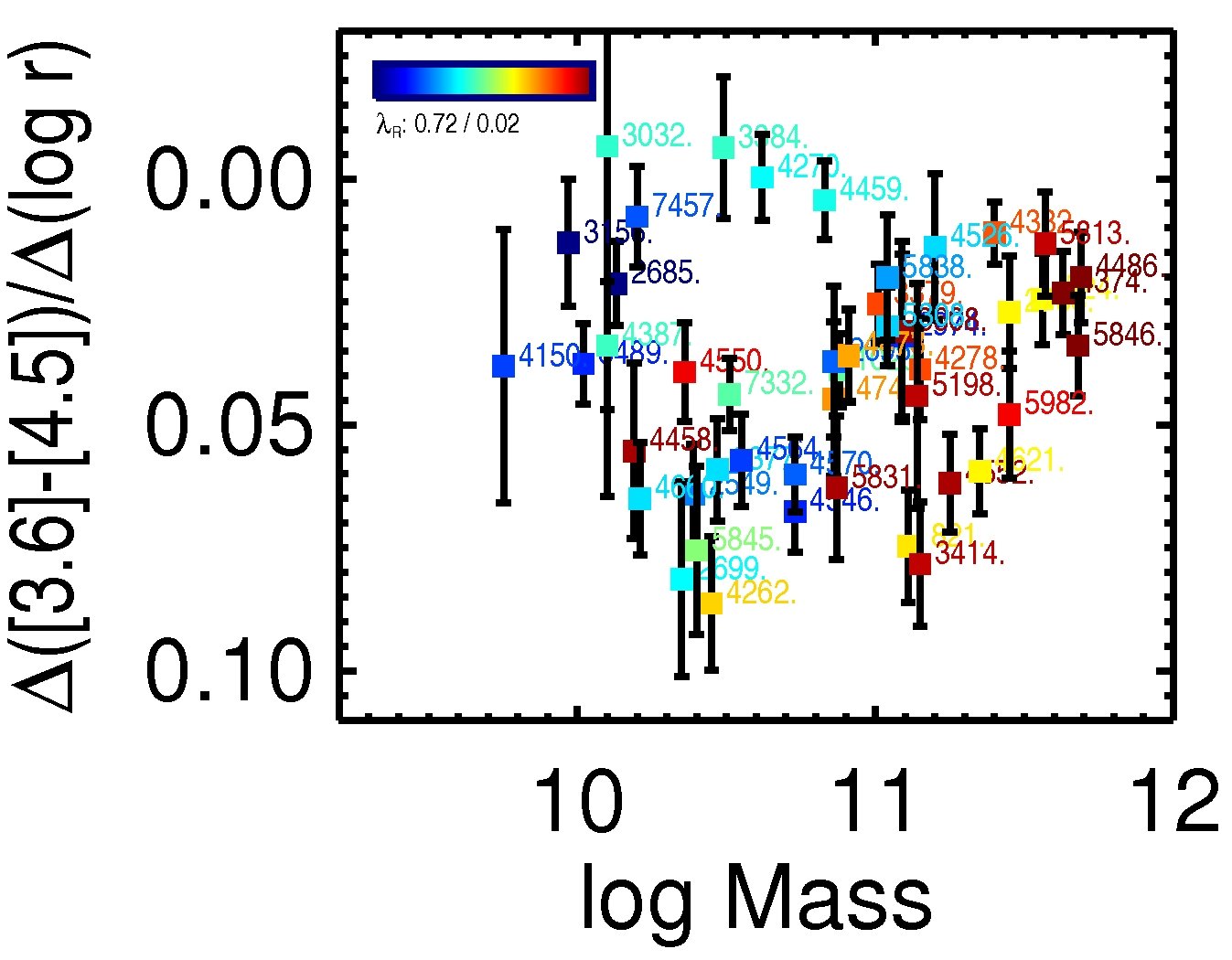}
\includegraphics[width=0.39\textwidth]{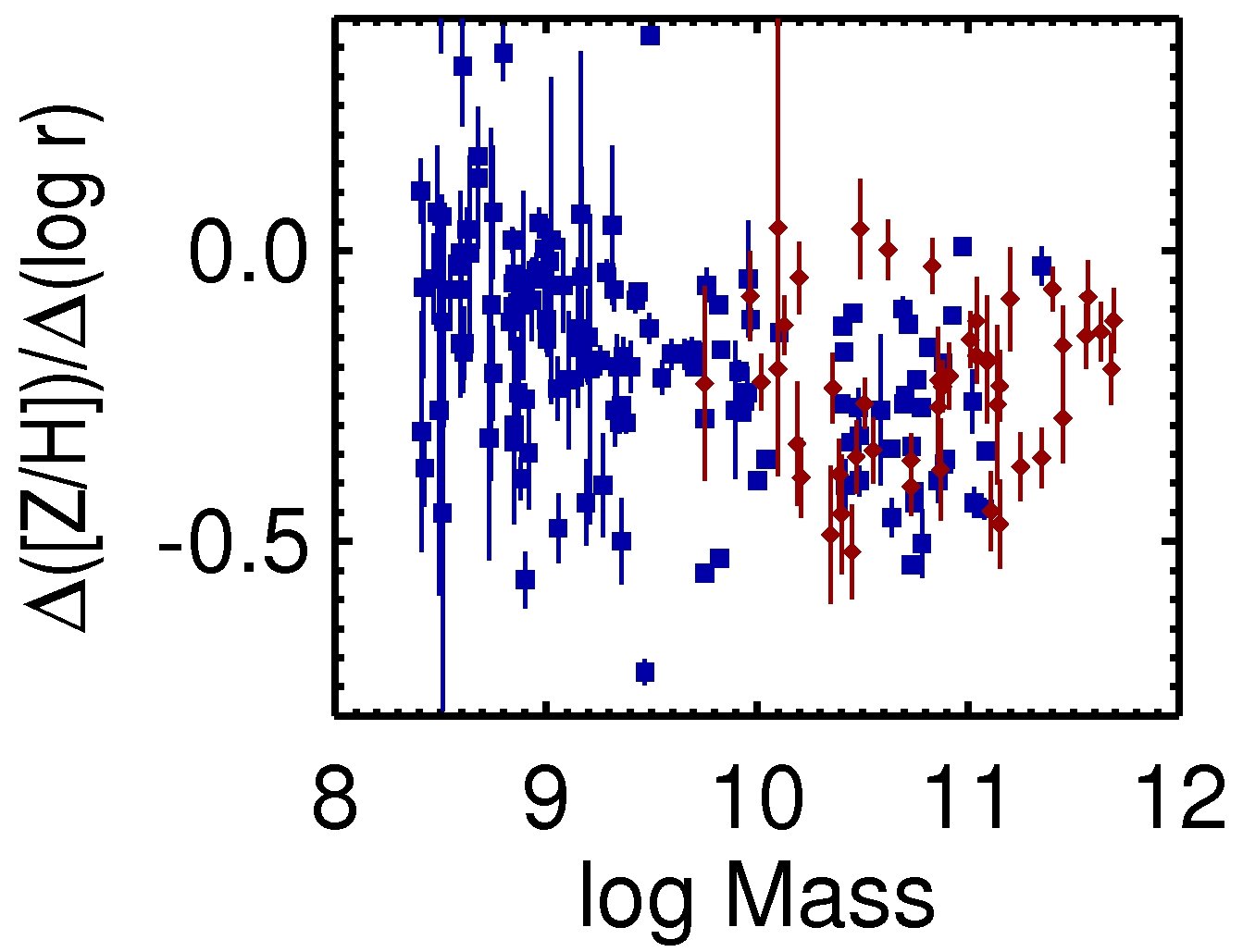}
\caption{
(a) and (b): $[3.6]-[4.5]$ colour gradient as a function of galaxy mass. In Fig.~\ref{colgrad_sigma}a 
the points have been coloured using the boxy/diskiness inside r$_e$ 
(from Paper IX). In Fig.~\ref{colgrad_sigma}b they have been coloured  using their $\lambda_R$ (Paper IX). 
(c) Colour gradient, converted to metallicity gradient, as a function of mass. 
Plotted here are the sample of this paper (red) and the Coma sample of den 
Brok et al. (2011, blue).  In Fig.\ref{colgrad_sigma}(a) and (b) dynamical masses are plotted. In 
Fig.\ref{colgrad_sigma}(c) the stellar masses of den Brok et al. (2011) are used, all in solar units. } 
\label{colgrad_sigma}
\end{center}
\end{figure}

In Fig.~\ref{colgrad_sigma}(a) b) we plot the [3.6] - [4.5] colour gradient as a
function of central mass.  We find that all gradients are positive within the
errors (indicating that metallicity decreases going outwards), and that there is
a small tendency for gradients to be smaller for less massive galaxies, although
the scatter at any $\sigma$ is large.  The scatter is clearly real, not due to
measurement errors. The good correlation with the metallicity gradients of Paper
XVII shows that the range in colour gradients is consistent with other papers in
the literature. Up to now we don't understand very well what determines the
metallicity gradient in a galaxy. For an in-depth discussion see den Brok et al.
(2011). It is a fact that metallicity gradients do not correlate with many
parameters, although several correlations are predicted. Classical monolithic
collapse scenarios (e.g. Carlberg 1984) predict large gradients, correlating
with galaxy mass. Such gradients can be diluted by mergers, although residual
star formation will again increase them in size (Hopkins et al. 2009).
Simulations are not very succesful in reproducing the observations, especially
since the observational results are not very clear. Another prediction is that
metallicity gradients ought to correlate with the orbital structure in the
galaxy. A cold disk has less mixing than a pressure supported elliptical galaxy.
For our sample, we have tried to test this by colouring the points using the
photometric boxy/diskiness parameter a$_4$/a (e.g. Carter 1978), and the
parameter $\lambda_R$ (Paper IX) measuring the rotational support of a galaxy.
Note that both parameters depend on the inclination of the galaxies, which is
not easy to determine. In Fig.~\ref{colgrad_sigma}(a) we find that boxy galaxies
possibly have smaller gradients than disky ones. If
a galaxy is boxy, orbit smearing will dilute the observed metallicity gradient.
Using this line of argument, however, metallicity gradients should also
depend on $\lambda_R$. A galaxy with a large $\lambda_R$ has a higher angular
momentum, and should have a larger gradient than a slow rotator. This, however,
is not observed (Fig.~\ref{colgrad_sigma}(b)). It might be that the sample of
slow rotators is too small. One should also realise that galaxies could have
different gradients in the radial and in the vertical directions, implying that
any range in inclination will smear out the gradients as well.

In Fig.~\ref{colgrad_sigma}(c) we expand the mass-range for the metallicity
gradient vs. mass relation by adding the Coma sample of den Brok et al. (2011).
The gradients there have been calculated in the same way, i.e. by excluding central
components, if those were visible in the colour profiles. We have converted the
data of den Brok et al. to metallicity gradients using  $\Delta(F475W - F814W) =
0.27 \Delta[Z/H]$, from that paper. We convert our $\Delta([3.6]-[4.5])$
gradients to metallicity multiplying them by a factor 5 (from
Fig.~\ref{metallicity_grad}). The figure shows that in the range of the bright
galaxies (our sample) the range in gradients in Coma galaxies is similar to the
range in the SAURON sample. Going then to fainter galaxies, the gradients slowly
become smaller. Note several dwarf galaxies with $10^8 M_\odot~ <$ Mass $< ~10^{10} M_\odot$ with very
large negative gradients. These are compact ellipticals (den Brok et al. 2011).
The qualitative behaviour of the galaxies in this diagram agrees with other
studies, e.g. S\'anchez-Bl\'azquez et al. (2006).

We have investigated whether these color gradients can be compared with predictions
from cold accretion models (e.g. Kere{\v s} et al. 2005, Brooks et al. 2009, Oser et al. 2010). Although
these models are becoming more and more accurate, they are not able to make predictions for metallicity 
gradients in galaxies. They do however predict that smaller galaxies contain a larger fraction of younger 
stars in their inner regions. This is indeed qualitatively what is observed, both in the global ages inside $r_e/2$
and in the very inner parts, where many of the smallest galaxies show young stars and ongoing star formation. 
We cannot use our observed relation between gradient and mass (Fig.~17), since here we have tried to 
give the gradients in the oldest stars, which are probably a measure of the metallicity gradients in our sample.

\subsection{Local colour and line strength vs. escape velocity}

\onecolumn
\begin{figure}
\centering
\includegraphics[width=0.98\textwidth]{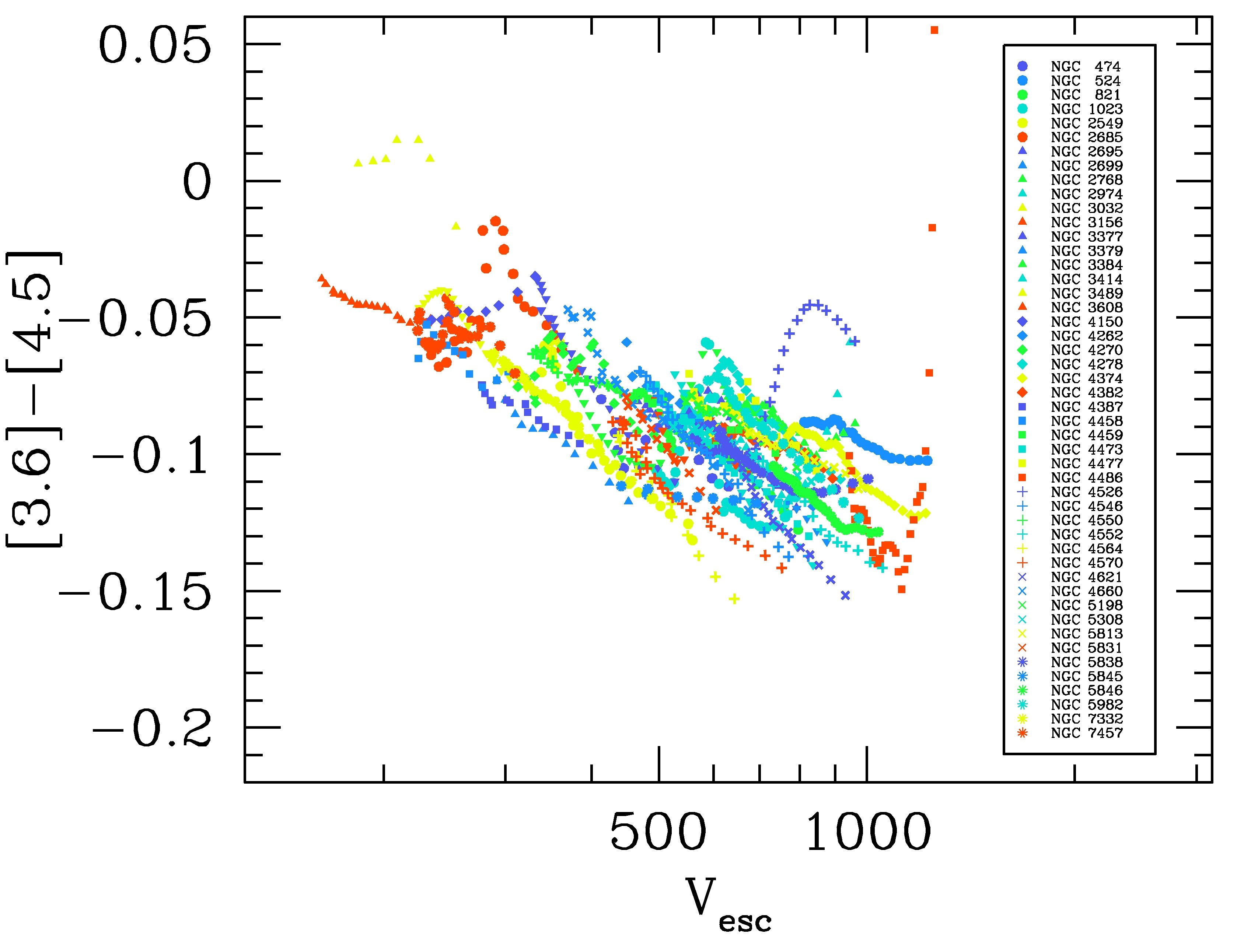}
\caption{Local colour  in mag~arcsec$^{-2}$ as a function of local escape velocity (in km/s). Plotted are all points of Table
\ref{tab_prof}. Escape velocities from Paper XIV).} 
\label{vesc}
\end{figure}
\twocolumn

In Scott et al. (2009, Paper XIV) we investigated the relation between the three
available line strengths and the local escape velocity for the SAURON sample,
prompted by the good correlation found by Franx \& Illingworth (1990) between
local colour and escape velocity. This dependence on a local parameter, which is
valid at all galaxy environments, and at different places inside a galaxy, might
suggest  that the gravitational potential, for which V$_{esc}$ is a proxy, is
the key parameter in determining the stellar populations in an early-type
galaxy.  Paper XIV found a tight relation between V$_{esc}$ and the line
strengths, especially Mg~{\it b}. For that index the relation from galaxy to
galaxy is the same as the relation within galaxies. For the other indices this
is not the case. Although Mg~{\it b} was found to be a very good proxy for
V$_{esc}$, there are several galaxies that do not fall on the Mg~{\it b} --
V$_{esc}$ relation, as a result of the sensitivity of Mg~{\it b} to young
stellar populations, which contribute relatively little to the mass. The fact
that for most galaxies the Fe 5015 -- V$_{esc}$ relation from galaxy to galaxy
is different than the relation within galaxies shows that, apart from the escape
velocity, the detailed metal enrichment history of a galaxy also plays a role in
determining the final stellar populations. It is therefore not surprising that
the Mg~{\it b} -- V$_{esc}$ relation has a measurable scatter. 

Here we investigate whether the [3.6] - [4.5] colour, which is less dependent on
young stellar populations, and correlates well with Mg~{\it b}, might show an
even tighter relation with escape velocity.  In Fig.~\ref{vesc} we show the
[3.6] - [4.5] colours presented in Fig.~4 as a function of escape velocity,
measured at each radius  (from Paper XIV).

There are a few qualitative differences visible between the [3.6] - [4.5] --
V$_{esc}$ and the Mg~{\it b} -- V$_{esc}$ relation. First, the young galaxies
now all are much closer to the mean [3.6] - [4.5] -- V$_{esc}$ relation. From
the 4 galaxies singled out in Paper XIV (NGC 3032, 3156, 4150 and 4382) only NGC
3032 still lies above the relation. This galaxy has a considerable young stellar
population (Paper VI, Paper XVII). Second, some central points of the galaxies
with AGN fall above the  [3.6] - [4.5] -- V$_{esc}$ relation. This can be seen
clearly for NGC 4486, and to a lesser extent for NGC 2768 and NGC 2974. Another
galaxy for which many points fall above the relation is NGC 4526. This galaxy
(see Fig.~\ref{n4526}),  which has a central dustlane, with young stars visible
from the H$\beta$ absorption line, interestingly enough does not fall above the
mean Mg~{\it b} -- V$_{esc}$ relation. It does, however, fall above the  [3.6] -
[4.5] -- V$_{esc}$ relation in the inner $\sim$ 10$''$, the region of the dust
lane, the region in which the [3.6] - [4.5] is redder than the extrapolated
outer gradient. This is the effect of young stellar populations (see Fig.~\ref{n4526}).  A
third remark about Fig.~\ref{vesc} is the fact that many galaxies that lie below
the [3.6] - [4.5] -- V$_{esc}$ relation also lie below the Mg~{\it b} --
V$_{esc}$ relation. This means that the scatter in both colours is not purely
caused by measurement errors (in agreement with the fact that the instrumental
errors in Mg~{\it b} are much smaller than the scatter in the Mg~{\it b} --
V$_{esc}$ relation (Paper XIV). For example, NGC 4458 and NGC 4387 lie below
both relations, while NGC 5845 and 2549 are completely above them. Detailed
stellar population analysis using more abosorption lines will be necessary to
show what is causing this effect. On top of this, one should realize that the
escape velocity depends on the assumed amount of dark matter in the galaxies, as shown
by e.g. Weijmans et al. (2009).  Also the slope of the galaxies in the colour --
$v_{esc}$ diagram changes. Since we don't know how much dark matter these
galaxies contain, we can only say that the small scatter from galaxy to galaxy
shows that the amount and distribution of the dark matter relative to the
visible light cannot be too different from galaxy to galaxy.

\begin{figure}
\centering
\includegraphics[width=0.49\textwidth]{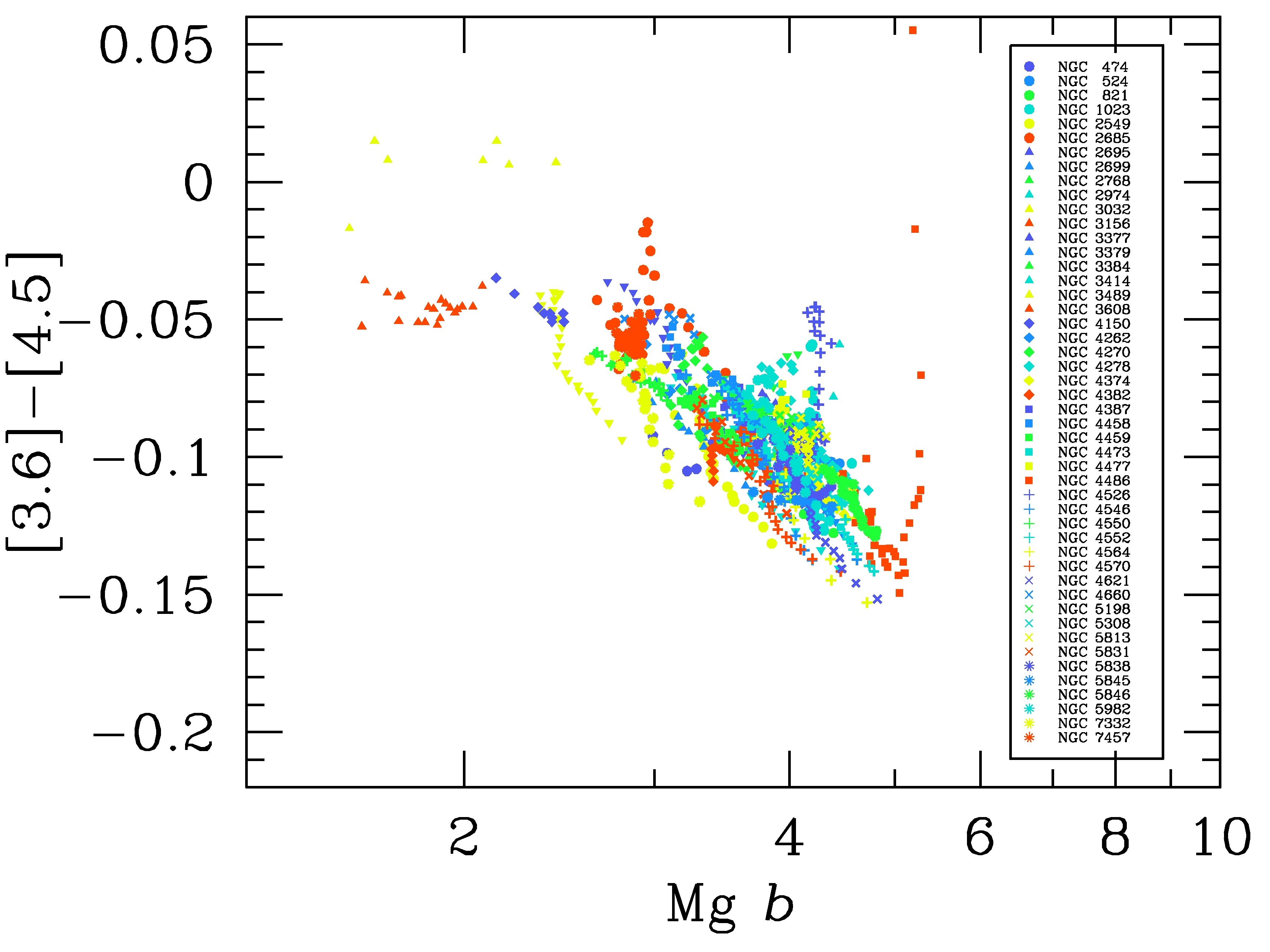}
\includegraphics[width=0.49\textwidth]{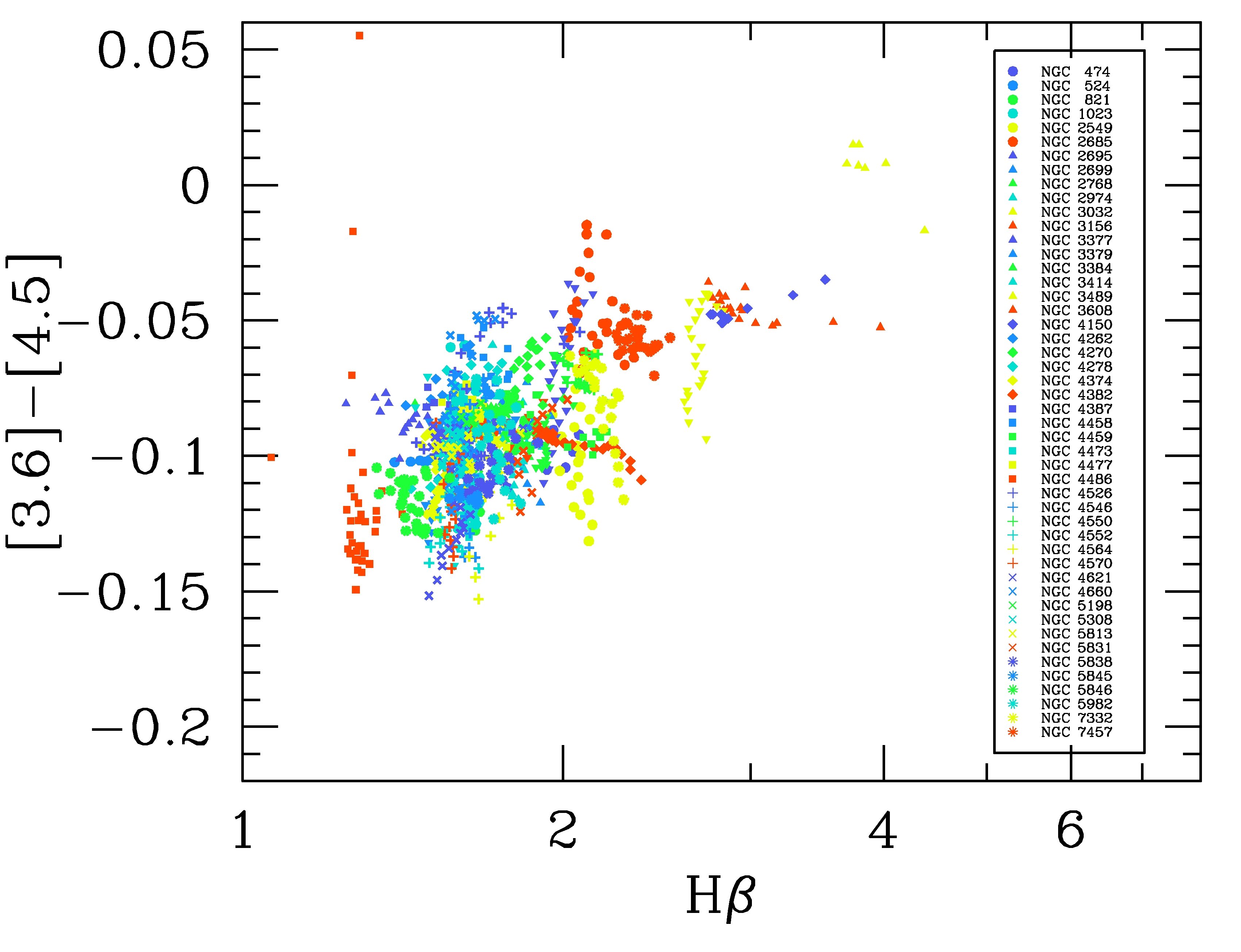}
\caption{Local colour  in mag~arcsec$^{-2}$  plotted as a function of local line strength index Mg~{\it b} (a) and H$\beta$
(b)  (both in \AA\ ) . Line indices from Paper VI.} 
\label{vesc_index}
\end{figure}

In Fig.~\ref{vesc_index} we show two local index-colour diagrams where the above
mentioned results can be seen again. Fig.~\ref{vesc_index}b shows that most
galaxies in the H$\beta$ vs. [3.6] - [4.5] diagram follow almost vertical
tracks. Within galaxies H$\beta$ is generally constant, while [3.6] - [4.5]
shows a small gradient. When considering integrated values, galaxies do,
however, show a reasonably strong relation between H$\beta$ and [3.6] - [4.5]
(see Fig.~\ref{indices}), consistent with Paper XIV.

\onecolumn
\begin{figure}
\begin{center}
\includegraphics{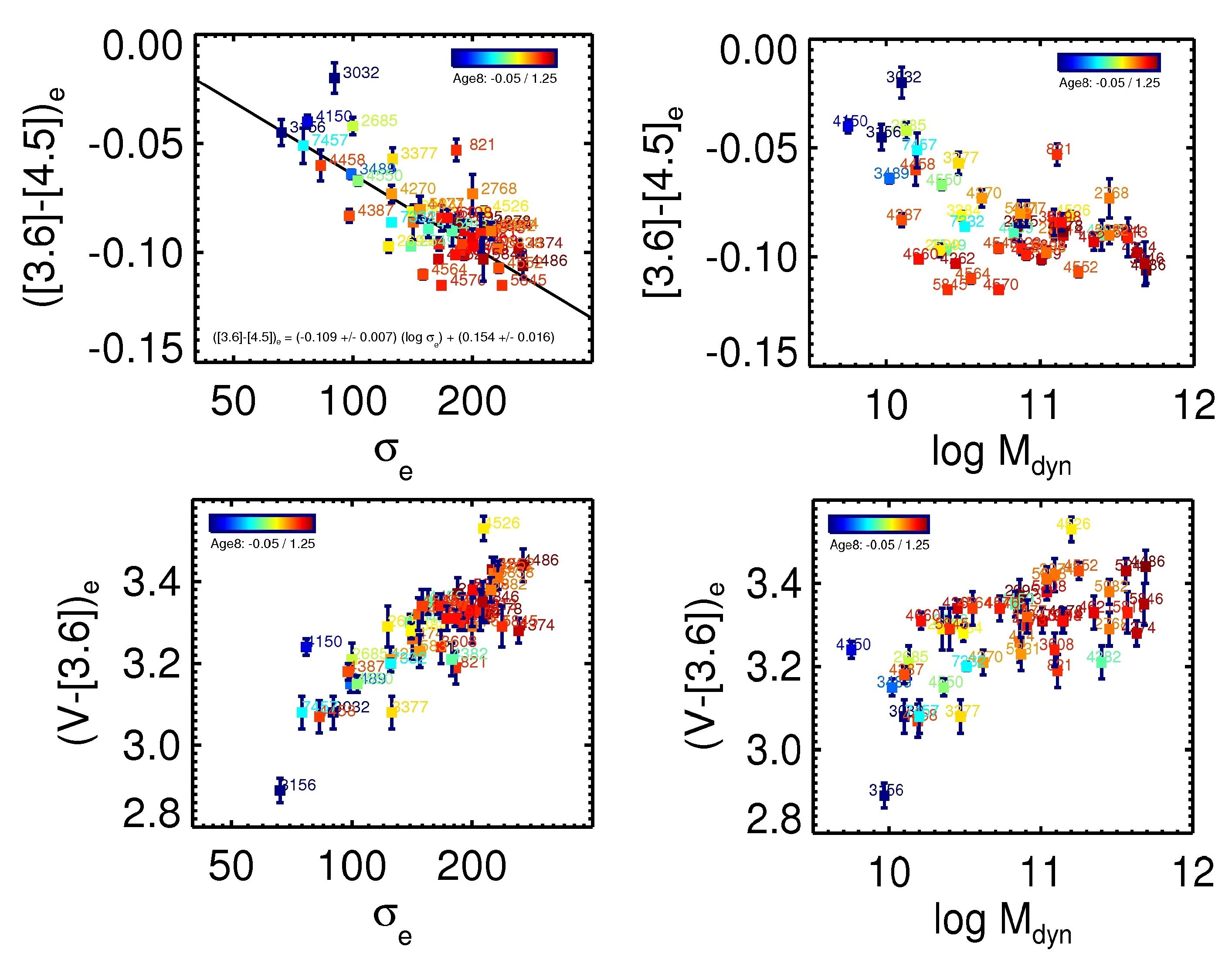}
\caption{Colour - mass (in solar mass) and colour - sigma (in km/s) relations for [3.6] - [4.5] and V - [3.6]. All colours have been calculated
within r$_e$.} 
\label{agemet}	 
\end{center}
\end{figure}
\twocolumn

\section{Discussion}

Recently, Temi et al. (2005) and Bregman et al. (2006) used spectra in the
mid-IR from 9-12 $\mu$m to establish that {\it Spitzer Space Telescope}
observations of elliptical galaxies are consistent with pure populations of very
old stars with no evidence of younger stars. Here we have used the IRAC [3.6] -
[4.5] colour, and show that  very accurate measurements can provide us with a
good metallicity indicator for the stellar populations, unaffected by dust
extinction, and that also young stellar populations can be detected. The fact
that this colour can be obtained relatively easily with a high S/N ratio offers
interesting possibilities for the future.

We find a tight $[3.6] - [4.5]$ vs. $\sigma$ relation. More massive galaxies
become bluer in this mid-infrared colour, contrary to other known optical and
near-infrared colours. The reason is that if the average temperature of the
stars decreases,  due to e.g., an increasing metallicity or a larger age, a
CO-band which dominates the 4.5$\mu$m  Spitzer filter becomes stronger, so that the
flux in the 4.5 $\mu$ filter decreases, and the  $[3.6] - [4.5]$ colour bluens,
despite the decreasing average temperature.   The scatter for the SAURON sample,
a representative sample of 48 ellipticals and S0 galaxies, is 0.015 mag, if one
takes the colour in apertures of $r_e$, and 0.024 mag, if a small aperture with
radius $r_e/8$ is used. Since the individual error in the data points is in
general smaller than 0.005 mag, this scatter is real. We show that the scatter
can be caused by hot dust near the center of a galaxy heated by its AGN, or by
the presence of young stellar populations, that are also seen in the H$\beta$
absorption line. Very few effects of AGB-stars are observed.

How to interpret the scaling relations? In Fig.~\ref{agemet} we have plotted [3.6] - [4.5]
as a function of $\sigma$ and mass (from Paper XIV). As mentioned before, the relation 
with mass is much more curved than the one with $\sigma$. For better understanding 
we also plot the same relations, but now having replaced [3.6] - [4.5] with V - [3.6],
a colour we understand better. In the relations of colour vs. mass, we see a common 
pattern that the oldest galaxies show the reddest colours in V - [3.6] and the bluest in 
[3.6] - [4.5]. For a fixed mass or $\sigma$ younger galaxies can be found at bluer colours 
in V - [3.6] and redder colours in [3.6] - [4.5]. 
When one compares the group of old galaxies that deviates from the straight [3.6] - [4.5] - mass 
relation (NGC 4570, 4564, 4262, 4660 and 5845) with other galaxies of the same mass, one
sees that their stellar populations are older, their $\sigma$ is larger, and their V - [3.6]
is redder. It looks as if stellar populations depend more on $\sigma$ then on total mass,
something which might be expected from the fact that local stellar population indicators 
correlate so well with escape velocity.
From this analysis we should conclude that the effects of metallicity on V - [3.6] for this sample are small, and the same probably also holds for [3.6] - [4.5] as a function of mass. What 
we are seeing is that the trends of V - [3.6] and [3.6] - [4.5] with mass and sigma are 
mostly due to age. Here one has to realize that this is SSP - age; a change in age means that 
the luminosity-weighted age is changing, by e.g. addition of a young component. 

We find that the scatter in the $[3.6] - [4.5]$ vs. $\sigma$ relation is
considerably larger when using the small aperture of $r_e/8$ than when using
$r_e$. We think that this is because at larger radii the stellar populations are
more uniform than in the central regions. Features are found in the $[3.6] -
[4.5]$ colour that correspond to central disks, rings, and bars, that are often
accompanied by young stellar populations. They are well visible in the line
strength maps of Paper VI and in the ionized gas emission maps of Paper V.
Although these young stellar populations, and also the red emission due to AGNs,
are the most obvious causes respondible for
the residuals in the $[3.6] - [4.5]$ vs. $\sigma$ relation, there might be
others, the most obvious one the contribution from evolved stars? Since most of the light in this infrared wavelength region comes from RGB and AGB stars, the contribution of AGB and RGB stars should vary as a function of age.  AGB stars would have deep 4.5 $\mu$m CO-bands, causing blue $[3.6] - [4.5]$ colours. The models of Marigo et al. (2008) confirm this. However, in younger galaxies we almost only see {\it redder} [3.6] - [4.5] colours. If a galaxy contains stars of 1-2 Gyr, which
do not show much H$\beta$ absorption any more, it still could contain many AGB
stars, making the [3.6] - [4.5] colour bluer for its H$\beta$ index. The fact that we don't see obvious cases where this is happening shows that we know very little about the behavior of the AGB phase in the mid-infrared. 

Some evidence for AGB populations might be found from absorption lines in the near-infrared.
M\'armol-Queralt\'o et al. (2009) measured  near-infrared line
indices in the K-filter in a number of ellipticals in the Fornax cluster and in
the field. Although the Mg~{\it b} - $\sigma$ relation is the same for both the
field and the Virgo sample, the CO index at 2.3 $\mu$m is much stronger (at the
same $\sigma$) in the field sample. In this paper it is argued that this is due
to an intermediate-age AGB population in the field. In the [3.6] - [4.5] vs. $\sigma$ relation, we find no difference between cluster and field galaxies, as defined in Paper II. This is not impossible, since the fraction of star-forming galaxies in
Fornax is much lower than in Virgo (e.g. Jord\'an et al. 2008), so that the likelihood for
intermediate-age stars in Virgo galaxies is also higher than in Fornax. 

The $[3.6]-[4.5]$ colour correlates well with a number of strong absorption
lines in the optical: Mg~{\it b}, Fe 5015 and H$\beta$. This is expected, since 
both the $[3.6]-[4.5]$ colour and the indices strongly depend on age and 
metallicity of the stellar populations, each in a different way.
These relations,
however, do also have some scatter that is not caused by observational
uncertainties. We have found that the [3.6] - [4.5] micron colours are redder
for galaxies with a compact radio source (AGNs) than for other galaxies, indicating some hot dust,
non-thermal emission, or simply young stars near the center. However, these younger stars 
are not detected in either Mg~{\it b}, Fe 5015 or H$\beta$. The effect in [3.6] - [4.5],
however, is so small, that we cannot say whether the reddening is due to hot dust or to young stars.
Some galaxies, showing completely old stellar populations, have bluer [3.6] - [4.5] colours
than other galaxies with old stellar populations. Here we might be seeing intermediate-age AGB stars.

Due to the technical difficulties in observing in this wavelength region, and
since stellar population studies are lacking here, the few theoretical stellar
population models available have barely been tested and used.  Since the models
of  Charlot \& Bruzual (2007)) give redder colours for increasing  metallicity, because they probably
do not take into account the strong CO-band at 4.5 $\mu$m, they cannot be used
to interpret these observations. The models of Marigo et al. (2008) on the other
hand do show the inverse behaviour observed in the data.  However, these models
also show that the $[3.6]-[4.5]$ colour is almost constant as a function of age,
which is probably not in agreement with the data. At this moment we can conclude
that larger, more massive, galaxies are more metal rich and bluer in
$[3.6]-[4.5]$. Young populations in the galaxies, primarily found in the inner
regions, make this colour redder locally.

We find that there is no simple correlation between color gradient and mass, not
even when the radial regions have been chosen in such a way that the color
gradient measures a metallicity gradient (Fig.~\ref{colgrad_sigma}). We do,
however, show that the [3.6] - [4.5] colour provides an excellent way to measure
metallicity gradients in many galaxies. For the bright galaxies of the SAURON
sample, we find that the colour gradients generally are positive, the
[3.6]-[4.5] colour becoming redder outwards. This implies, consistent with
studies in the optical and near-infrared $K$-filter, that galaxies are
less metal rich when going outwards. The comparison with the metallicity
gradients (from the optical, Paper XVII) is excellent, apart from a few galaxies
for which there is circumstantial evidence that the [3.6] -- [4.5] gradients are
also due to gradients in age. We find a hint that gradients for boxy galaxies are
smaller on the average than those for disky ones. This can be expected if boxy
galaxies have a larger fraction of low angular momentum orbits, which will
dilute the three-dimensional gradients. On the other, if this were the case, one
would also expect pressure-supported galaxies with low $\lambda_R$ values to
have smaller gradients than those with high values of $\lambda_R$. This effect
is possibly too difficult to measure for our sample, given the dispersion caused
by the range in inclination. A larger sample, e.g. the ATLAS$^{\rm 3D}$ sample
(Cappellari et al. 2011), would be needed to establish the dependence of
metallicity gradient on orbital structure.

\section{Conclusions}
\label{sec:conc}

In this paper, we have studied the Spitzer-IRAC $[3.6] - [4.5]$ colour, which for early-type galaxies is dominated by emission from stellar photospheres, is independent of dust extinction, and can be measured easily down to low
surface brightness levels on the Spitzer data.

Here we summarize our main results. 

\begin{itemize}
\item The Spitzer-IRAC $[3.6] - [4.5]$ colour can be used very well to study the stellar populations in early-type galaxies.
\item Local early-type galaxies display a tight colour-velocity dispersion relation, with more
massive galaxies showing bluer colours. 
\item Although we also find tight relations of [3.6] - [4.5] with mass and 3.6 $\mu$m luminosity, 
the latter are more curved than the one with $\sigma$. This has to do with the mass distribution 
within the galaxies and the change of slope in the Faber-Jackson relation.
It looks as if stellar populations depend more on $\sigma$ then on total mass.
\item Deviations from the colour - $\sigma$ relations, larger than the observational
uncertainties, are seen. Several of them are identified to be due to young stellar populations
detected in the line strength maps of Kuntschner et al. (2006), in CO, in ionized gas, in UV and at 8 $\mu$m, 
which cause the $[3.6] - [4.5]$ colour to redden. 
\item In the galaxies that contain compact radio sources, as detected by the 
VLA FIRST survey, the [3.6] - [4.5] colour is slightly redder than in other galaxies of the same mass. This
is due either to hot dust near the AGN, non-thermal emission, or young stars in the very center of these galaxies.
\item To be able to interpret the data well, stellar population models need to be improved.
However, we can already say that the $[3.6] - [4.5]$ colour seems to be a good
metallicity indicator, with this colour becoming bluer with increasing metallicity. Better
modelling will have to show whether we can use the $[3.6] - [4.5]$ colour together with
an optical line index to separate age and metallicity in galaxies. 
\item We obtained radial colour profiles and measured colour gradients
$\Delta([3.6]-[4.5])/\Delta \log R$ using the (outer) part of the radial colour profile that can be
represented by a single linear fit, indicating a change in metallicity. Here   we investigated the relation between
colour gradients of old stellar populations (metallicity gradients) and galaxy mass, luminosity and $\sigma$.  We find that colour gradients are generally positive, indicating
that galaxies become redder going outwards, or slighty more metal poor.
\item The spread in gradients, especially for more
massive galaxies, is real, and much larger than can be accounted for by observational uncertainties.
\item The local [3.6]-[4.5] colour shows a  correlation with local escape velocity that is
possibly even tighter than Mg~{\it b}, since the dependence on age is smaller. The fact that
this relation exists shows that the dark matter distribution, as compared to the luminous matter,
is very similar from galaxy to galaxy.
\end{itemize}

The results presented in this paper show that the [3.6] - [4.5] colour is a useful stellar population measure, which is relatively easy to measure with high S/N in galaxies, when one has space-based data. In this paper we have shown that it is sensitive both to  metallicity and Star Formation History. Since spectra in this region of the spectrum are extremely scarce, our understanding of the behaviour of this colour is limited. Moreover, available stellar population models are rather uncertain since evolved stages of stellar evolution are difficult to model. We will need to wait until the launch of the James Webb Space Telescope (JWST), for this situation to improve. At that time we get spectra in this wavelength region that will show what lies behind the scaling relations shown in this paper.

\section*{Acknowledgments} We thank Scott Trager for fruitful discussions, and Sandro Bressan for some very
useful suggestions. This work is based on observations made with the {\it Spitzer Space Telescope}, which is
operated by the Jet Propulsion Laboratory, California Institute of Technology under a contract with NASA.
Support for this work was provided by NASA through an award issued by JPL/Caltech. 
The SAURON project was made possible through grants 
from NWO and financial contributions from the Institut National des Sciences de l'Univers, the Universit\'e
Lyon I, the Universities of Durham, Groningen, Leiden and Oxford, the Programme National Galaxies, the British
Council, PPARC grant {\it Observational Astrophysics at Oxford 2002-2006} and support fromSpace
Christ Church Oxford, and the Netherlands Research School for Astronomy NOVA. 
GvdV acknowledges support provided
by NASA through Hubble Fellowship grant HST-HF-01202.01-A awarded by the {\it Space Telescope Science Institute},
which is operated by the Association of Universities for Research in Astronomy, Inc., for NASA, under
contract NAS 5-26555. JFB acknowledges support from the Ram\'on y Cajal Program as well as grant AYA2010-21322-C03-02 by the Spanish Ministry of Science and Innovation. This paper is also based on observations obtained at the William Herschel Telescope,
operated by the Isaac Newton Group in the Spanish Observatorio del Roque de los Muchachos of the Instituto
de Astrofsica de Canarias. This project made use of the
NED database. MC acknowledges support from a Royal Society University Research Fellowship.


\end{document}